\documentclass[useAMS,usenatbib]{mn2e}
\usepackage{graphicx}
\usepackage{subfigure}
\usepackage{epsfig}
\usepackage{amsmath}
\usepackage{natbib}
\usepackage{color}
\usepackage{setspace}

\def\kms{km\,s$^{-1}$}

\title[Non-thermal excitation and ionization in supernovae]{Non-thermal
excitation and ionization in supernovae}
\author[Chengdong Li, D. John Hillier $\&$ Luc Dessart]
{Chengdong Li$^{1}$, D. John Hillier$^{1}$ $\&$ Luc Dessart$^{2}$\\
$^{1}$Department of Physics and Astronomy $\&$ Pittsburgh Particle physics, Astrophysics, and Cosmology Center (PITT PACC), \\University of Pittsburgh, Pittsburgh, PA 15260, USA\\
$^{2}$Laboratoire d'Astrophysique de Marseille, Universit\'e Aix-Marseille $\&$ CNRS, UMR7326, 38 rue Fr\'ed\'eric Joliot-Curie, \\13388 Marseille, France }

\begin{document}

\date{\today}

\pagerange{\pageref{firstpage}--\pageref{lastpage}} \pubyear{2012}

\maketitle
\label{firstpage}

\begin{abstract}
We incorporate non-thermal excitation and ionization processes arising from non-thermal electrons  that result from $\gamma$-ray energy deposition, into our radiative transfer code {\sc cmfgen}. The non-thermal electron distribution is obtained by solving the Spencer-Fano equation using the procedure of \cite{1992ApJ...390..602K}. We applied the non-thermal calculations to the blue supergiant explosion model whose early evolution was studied in \citet{2010MNRAS.405.2141D}. Non-thermal processes generally increase excitation and ionization and decrease the temperature of the ejecta. We confirm that non-thermal processes are crucial for modeling the nebular spectra. Both optical H\,{\sc i} and He\,{\sc i} lines are significantly strengthened. While optical He\,{\sc i} lines are not easily discerned in observational spectra due to severe blending with other lines, He\,{\sc i}\,2.058\,$\mu$m provides an excellent opportunity to infer the influence of non-thermal processes. We also discuss the processes controlling the formation of the He\,{\sc i} lines during the nebular epoch. Most lines of other species are only slightly affected. We also show that the inclusion of Fe\,{\sc i} has substantial line-blanketing effects on the optical spectra. Our model spectra and synthetic light curves are compared to the observations of SN 1987A. The spectral evolution shows broad agreement with the observations, especially H$\alpha$. The uncertainties of the non-thermal solver are studied, and are expected to be small. With this new addition of non-thermal effects in {\sc cmfgen}, we now treat all known important processes controlling the radiative transfer of a supernova ejecta, whatever the type and the epoch.
\end{abstract}


\section{INTRODUCTION}
\label{intro}

Supernovae (SNe) explosions are classified as thermonuclear SNe and Core-Collapse SNe (CCSNe), according to their explosive mechanisms. Thermonuclear SNe are also called Type Ia SNe, with light curves that can be standardized to measure cosmological distances. Based on their spectroscopic signatures, CCSNe can be further categorized into 3 broad classes, Type Ib, Type Ic and Type II. Type II SNe exhibit hydrogen lines, while Type I SNe do not.

SNe explosively produce $^{56}\textrm{Ni}$ which decays into $^{56}\textrm{Co}$ with a half-life of 6.1 days. $^{56}\textrm{Co}$ further decays into $^{56}\textrm{Fe}$ with a half-life of 77 days. Such radioactive decay is the main energy source of Type Ia SNe. While the energy source of Type Ib/c SNe resembles that of Type Ia SNe, Type II SNe are first powered by the shock-deposited energy, but at later times radioactive decay will be the dominant energy source. High-energy $\gamma$-rays are produced during the radioactive decay. The $\gamma$-rays undergo Compton scattering with bound and free electrons, which results in high-energy electrons. These high-energy electrons interact with the medium and deposit energy through heating, excitation, and ionization.

Non-thermal excitation and ionization processes have been suggested to explain some observational features of various SNe of different types. For example, the Bochum event in the H$\alpha$ profiles of SN 1987A was explained by non-thermal excitation due to an asymmetric ejection of $^{56}\textrm{Ni}$ \citep{1989PASP..101..137P,1991SvAL...17..400C}. \citet{1995ApJ...441..821L} included two-photon processes in models for SN 1987A, and showed that they could fit the evolution of the He\,{\sc i}\,1.083\,$\mu$m and 2.058\,$\mu$m lines whose upper levels are populated by non-thermal processes induced by $\gamma$-rays.

Several groups \citep{1987ApJ...317..355H,1991ApJ...383..308L,1991ApJ...373..604S, 1992ApJ...390..602K, 1993ApJ...411..313S,1998ApJ...496..946K,1998ApJ...497..431K} studied He\,{\sc i} lines in Type Ib SNe and suggested these lines are due to non-thermal excitation. \citet{1987ApJ...317..355H} carried out LTE calculations for Type Ib SNe and showed that these did not match observations -- emission line strengths indicate large departures from LTE for He\,{\sc i}. \citet{1991ApJ...383..308L} treated non-thermal excitation and ionization for He\,{\sc i} in their LTE models, and found that it could explain the presence and strength of He\,{\sc i} lines. However, these studies did not treat all atoms in non-LTE and/or ignored effects due to radiative transfer. Recently, \cite{2011MNRAS.414.2985D} carried out a comprehensive study of SNe which result from quasi-hydrogen-less low-mass cores from 15-25~M$_{\odot}$ main-sequence stars in binary systems using {\sc cmfgen}. They found that the He\,{\sc i} lines could be reproduced at early times if the He mass fraction is greater than 50 per cent, without invoking non-thermal excitation.

Line formation in SNe is inherently complex. The fast expansion velocities exacerbate the influence of lines, by producing broader lines, increasing line overlap, and enhancing line blanketing. Line blanketing strongly influences the optical and UV ranges of SNe spectra, generally producing a flux deficit in those regions. To accurately account for the  influence of line blanketing requires that an enormous number of metal lines be included in the model. Because of non-LTE and time dependence, an accurate interpretation of the spectral signatures requires quantitative studies.

For our studies we utilized {\sc cmfgen}, which is a radiative transfer code developed for modeling Wolf-Rayet stars \citep{1987ApJS...63..947H} that was later enhanced to allow for line blanketing \citep{1998ApJ...496..407H}, to model all early-type stars, and ultimately to model all types of SNe \citep{2012arXiv1204.0527H}. The code includes the following processes: bound-free; bound-bound (including two-photon emission); free-free; dielectronic recombination; electron scattering; Rayleigh scattering by hydrogen; inner-shell ionization by X-rays; collisional ionization/recombination; collisional excitation/de-excitation and charge exchange reactions with H and He. The code does full-non-LTE calculations at great computational cost. Atomic data utilized
by {\sc cmfgen} are listed by \cite{2010MNRAS.405.2141D}.

Time dependent terms in the rate equations are usually neglected in SNe spectral modeling at early times, because the recombination time is much smaller than the age of the SN. However, \citet{2005A&A...441..271U} showed that time-dependent ionization of hydrogen is responsible for reproducing the strong H$\alpha$ lines at early times in SN 1987A, which was confirmed by \citet{2008MNRAS.383...57D}. {\sc cmfgen} takes into account the time-dependent terms in the statistical and radiative equilibrium equations, and in the radiative transfer equations. The code has been extensively tested and used to study various types of SNe \citep{2005A&A...437..667D,2006A&A...447..691D, 2008MNRAS.383...57D,2010MNRAS.405.2141D,2011MNRAS.410.1739D, 2008ApJ...675..644D,2012arXiv1204.0527H}.

At early times, non-thermal effects in Type II-P SNe are small, since the optical depth is large and the ejecta is essentially in LTE in the region where $^{56}\textrm{Ni}$ is abundant. However, non-thermal effects will eventually take over other processes, especially during the nebular phase. Further, the study of \cite{2011A&A...530A..45J} shows that radiative transfer effects are still very important at very late times. In Type Ib SNe, non-thermal processes can become important as early as 10 days \citep[e.g.,][]{1991ApJ...383..308L,2011MNRAS.414.2985D,2012inprep0000.0001D}. Thus non-thermal excitation and ionization, as well as radiative transfer effects, need to be taken into account in order to undertake comprehensive studies of SNe from the photospheric through the nebular phase.

With the implementation of non-thermal processes in {\sc cmfgen}, we are now able to model from just after the explosion, through the photospheric phase, the onset of the nebular phase, and the nebular phase using a single code. The aim of this paper is to describe the implementation of non-thermal excitation and ionization into {\sc cmfgen}, and to benchmark the non-thermal solver using SN 1987A. The paper is organized as follows: We present our methodology of the non-thermal solver in Section \ref{sec:method}, and the hydrodynamical model, utilized in our studies, is summarized in Section \ref{sec:hydro}. Non-thermal effects are discussed in Section \ref{sec:ntmodel}, while we compare the properties of the non-thermal and thermal models in Section \ref{sec:mod_cmp}. To examine the importance of additional opacity sources we study the influence of Fe\,{\sc i} in Section \ref{sec:fei}. Comparison to the observations is made in Section \ref{sec:obs_cmp}. We explore the uncertainties underlying the non-thermal solver in Section \ref{sec:uncertainties}, and discuss the effect of mixing and $\gamma$-ray transport in Section \ref{sec:discussion}. Finally, we will summarize our results in Section \ref{sec:conclusion}. A parallel paper will discuss the influence of non-thermal processes on Type Ibc spectra \citep{2012inprep0000.0001D}.

\section{The Method}
\label{sec:method}

Non-thermal excitation and ionization enter the radiative transfer calculations by changing the level populations. To quantify these excitation and ionization rates, the degradation spectrum, which describes the number of fast electrons as a function of energy, needs to be known. There are several ways to solve for the degradation spectrum \citep{1954PhRv...93.1172S,1985ApJ...298..268S,1989ApJ...343..323F,1991ApJ...375..190X,1992ApJ...390..602K}. One approach is the Monte Carlo method which, in general, is slow. Our approach is to solve the Spencer-Fano equation following the method of \citet{1992ApJ...390..602K}. Once the degradation spectrum is known, all non-thermal excitation and ionization rates are computed and included in the statistical equilibrium equations (SEE).

\subsection{The Spencer-Fano Equation}
\label{sec:sf_eq}

The radiative transfer code {\sc cmfgen}, which takes into account radioactive decay, previously assumed that all radioactive decay energy is deposited as heat. In reality, fast electrons resulting from Compton scattering between $\gamma$-ray and thermal electrons lose energy through three different channels -- heating, excitation and ionization. The fractions of energy for the three channels can be computed if the degradation spectrum is known. This can be done by solving the Spencer-Fano equation \citep{1954PhRv...93.1172S}. Following \citet{1992ApJ...390..602K}, the Spencer-Fano equation, which balances the number of electrons leaving and entering an energy interval, can be written as:

{\footnotesize
\begin{equation}\label{eq:spencer}
\begin{aligned}
& \sum_i n_i y(E) \left[  \sum_{j} \sigma_{ij}(E)
+ \int_{I}^{(I+E)/2} \sigma_{c}(E,\epsilon) d\epsilon  \right] \\[-1mm]
& - \frac{d}{dE} \left[ y(E)L_e(E) \right]
= \sum_i n_i  \sum_j y(E+E_{ij})\sigma_{ij}(E+E_{ij}) \\[-1mm]
& + \sum_i n_i \int_{I}^{\lambda} y(E+\epsilon)
\sigma_c(E+\epsilon,\epsilon)d\epsilon  \\[-1mm]
& + \sum_i n_i \int_{2E+I}^{E_{\textrm{max}}} y(E')\sigma_c(E',E+I)dE' + S(E)
\end{aligned}
\end{equation}
}

\noindent where $\lambda = \textrm{min} (E_{\textrm{max}}-E,E+I)$, $E_{\textrm{max}}$ is the maximum energy of non-thermal electrons, and $I$ is the ionization potential. $y(E)dE$ is defined as the flux of non-thermal electrons in the energy interval $(E,E+dE)$. $n_i$ is the number density of a single species, and the sum over $i$ allows for the contribution of different ionization stages and species. $L_e(E)$ is the energy loss of the non-thermal electrons due to Coulomb scattering by the thermal electrons. $E_j$ is the excitation energy, $\sigma_j$ is the excitation cross section of level $j$ and $\sigma_c(E,\epsilon)$ is the differential ionization cross section for a fast electron with energy $E$ and an energy loss of $\epsilon$. Therefore, the ejected electron has energy $\epsilon-I$. The fast electron is called the primary electron and the ejected electron is called the secondary electron. The secondary electron can further degrade through the three same channels as the primary electron. $S(E)$ is the source term, and we assume all non-thermal electrons are injected at $E_{\textrm{max}}$. The degradation spectrum $y(E)$ is normalized to $E_{\textrm{max}}$, so that $y(E)$ is the degradation spectrum for 1 eV energy input. The first and second term on the left side represent the number of electrons leaving the energy interval $(E,E+dE)$ by impact excitation and ionization, respectively. The term with $L_e(E)$ is the number of electrons leaving the energy interval by heating the medium. The first term on the right side is the number of electrons entering the energy interval $(E,E+dE)$ by exciting another ion and losing energy. The second term is the ionization contribution from the primary electrons, and the third term is the ionization contribution from the secondary electrons if applicable.

The Spencer-Fano equation is more conveniently solved numerically using the integral form \citep{1992ApJ...390..602K}, i.e.,

{\footnotesize
\begin{equation}\label{eq:spencer_int}
\begin{aligned}
\sum_i & n_i \sum_j \int_{E}^{E+E_j} y(E') \sigma_j (E')dE'
+  y(E) L_e(E)  \\[-1mm]
& + \sum_i n_i \int_{E}^{E_{\textrm{max}}} y(E') \int_{E'-E}^{(E'+E)/2}
\sigma_c(E',\epsilon) d\epsilon dE' \\[-1mm]
& = \sum_i n_i \int_{2E+I}^{E_{\textrm{max}}} y(E')
\int_{E+I}^{(E'+I)/2}\sigma_c(E',\epsilon) d\epsilon dE' \\[-1mm]
& + \int_{E}^{E_{\textrm{max}}} S(E')dE'
\end{aligned}
\end{equation}
}

\noindent In Eq.~\ref{eq:spencer}, the integral of the differential ionization cross sections needs to be done numerically. However, only the total ionization cross sections appear in the integral form of the Spencer-Fano equation (Eq.~\ref{eq:spencer_int}), and can be computed analytically.

With the solution of the degradation spectrum, the fractional energies entering heating, $\eta_h$, excitation of ion $i$ to level $j$, $\eta_{ij}$, and ionization of ion $i$, $\eta_{ic}$, can be computed by

\begin{equation}
\begin{aligned}
\eta_h & = \frac{1}{E_{\textrm{init}}} \int_{E_0}^{E_{\textrm{max}}} y(E')L_e(E')dE' \\
& + \frac{1}{E_{\textrm{init}}} E_0 y(E_0)L_e(E_0)
+ \frac{1}{E_{\textrm{init}}} \int_0^{E_0} N(E')E'dE'
\end{aligned}
\end{equation}

\begin{equation}
\eta_{ij} = \frac{n_{ij}E_{ij}}{E_{\textrm{init}}} \int_{E_{ij}}^{E_{\textrm{max}}}
y(E')\sigma_{ij}(E')dE'
\end{equation}

\begin{equation}
\eta_{ic} = \frac{n_i I_i}{E_{\textrm{init}}} \int_{I_{i}}^{E_{\textrm{max}}}
y(E')\sigma_{ic}(E')dE'
\end{equation}

\noindent where

\begin{equation}
\begin{aligned}
N(E) & = \sum_i n_i \left[ \sum_j y(E+E_{ij})\sigma_{ij}(E+E_{ij}) \right. \\
& + \int_{I_i}^{\lambda_i} y(E+E')\sigma_{ic}(E+E',E')dE' \\
& + \left. \int_{2E+I_i}^{E_{\textrm{max}}} y(E')\sigma_{ic}(E',E+I_i)dE' \right] + S(E)
\end{aligned}
\end{equation}

\noindent \citep{1992ApJ...390..602K}. $E_{\textrm{init}}$ is the mean energy of the initial electrons, and $E_0$ is the lowest excitation or ionization energy.

To save computational effort, we only take into account the dominant ionization stages when computing the degradation spectrum. The other approximation is the neglect of $\gamma$-ray transport. In reality, some $\gamma$-rays may escape the SN, while others may deposit their energy at large distances from where they were created. In these calculations we assume, for simplicity, in-situ deposition. A $\gamma$-ray transport code has been developed for use with {\sc cmfgen} \citep{2012arXiv1204.0527H}.

We previously mentioned that for the source term we assume that all high energy electrons are injected at $E_{\textrm{max}}$. Although not entirely realistic, this kind of source function generally has little influence on the results. High energy electrons ``forget'' their initial energies after several scatterings, and the source function is quickly washed out. Alternatively we can assume a bell shape source function near $E_{\textrm{max}}$. This tends to give a smoother degradation spectrum near $E_{\textrm{max}}$.

\subsection{Ionization cross sections}
\label{sec:arnaud}

\cite{1985A&AS...60..425A} adopted the formula proposed by \cite{1981JQSRT..26..329Y} for calculating the impact ionization cross sections

\begin{equation}\label{eq:arnaud}
\begin{aligned}
Q_i(E)&  = \frac{1}{uI_i^2} \{  A_i \left(1-\frac{1}{u}\right)
+ B_i \left( 1-\frac{1}{u} \right)^2 + \\
& + C_i \ln (u) + D_i \frac{\ln(u)}{u} \}
\end{aligned}
\end{equation}

\noindent where $u = E/I_i$. E is the energy of the impact electron, and $I_i$ is the ionization potential for the element. $A_i$, $B_i$, $C_i$, and $D_i$ are all coefficients. For the meaning of these coefficients, the reader should refer to \cite{1985A&AS...60..425A}. In our calculations, we have used the most up-to-date coefficients (Arnaud, private communication).

With the total cross section computed by Eq.~\ref{eq:arnaud}, the differential cross section $\sigma_c (E,\epsilon)$ can be written as,

\begin{equation}
\sigma_c(E,\epsilon) = Q(E) P(E,\epsilon - I)
\end{equation}

\noindent where $P(E_p,E_s)$ is the distribution of the secondary electrons for a primary electron energy of $E_p$. \citet{1971JChPh..55.4100O} found $P(E_p,E_s)$ could be well described by

\begin{equation}
P(E_p,E_s) = \frac{1}{J \arctan [(E_p - I)/2J ]} \frac{1}{[1+(E_s/J)^2]}
\end{equation}

\noindent where $J$ is a parameter that gives a best fit to measurements and varies for different species. We adopt $J = 0.6 I$, which is shown to be a reasonable approximation \citep{1971JChPh..55.4100O}. The secondary distribution function drops quickly so that very few secondary electrons have high energies. Specifically, for H\,{\sc i} with $I = 13.6$\,eV, 66.7\% of secondary electrons are below the ionization potential if $E_p = 1000$\,eV and 69.8\% in the case of $E_p = $10\,000\,eV. Further, the mean energy of the secondary electrons, $\bar{v}_{\textrm{sec}}$, only varies slowly as the energy of the primary electron $E_p$ increases. For hydrogen, $\bar{v}_{\textrm{sec}} = 21.53$\,eV for $E_p = 1000$\,eV and $\bar{v}_{\textrm{sec}} = 33.36$\,eV for $E_p = $10\,000\,eV. Secondary electrons with low energies are easily thermalized and deposit energy as heat.

\subsection{Excitation cross sections}
\label{sec:bethe}

\begin{figure}
  \includegraphics[bb=45 240 590 540, width=0.48\textwidth]
  {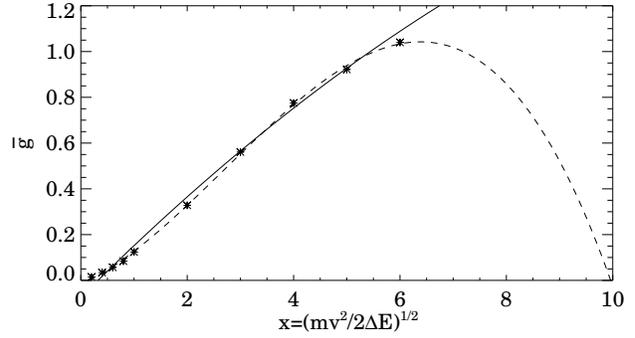}
  \caption{The tabulated gaunt factor (stars) for neutral atoms as a function
  of $x=\sqrt{\frac{mv^2}{2\Delta E}}$ given in \citet{1962ApJ...136..906V}. The
  second order (solid) and the third order (dashed) polynomial fit are also shown.}
  \label{fig:gaunt_fac_fit}
\end{figure}

The electron impact excitation cross sections are only available for a few species, and only for a few levels. To proceed further, we adopt the Bethe approximation as discussed by  \cite{1962ApJ...136..906V}. With this approximation, $Q_{ij}$ is given by

\begin{equation}
\begin{aligned}
Q_{ij} = \frac{8\pi}{\sqrt{3}}\frac{1}{k_i^2} \frac{I_H}{E_j-E_i}
f_{ij} \bar{g} \pi a_0^2
\end{aligned}
\end{equation}

\noindent where $k_i^2$ is the initial electron kinetic energy normalized to hydrogen ionization energy 13.6\,eV, $I_H$ is the hydrogen ionization energy, $E_j-E_i$ is the energy difference between the two levels, $f_{ij}$ is the oscillator strength, and $a_0$ is the Bohr radius. The Gaunt factor $\bar{g}$ for neutral atoms utilizes Table~1 of \cite{1962ApJ...136..906V}, which is shown in Fig.~\ref{fig:gaunt_fac_fit}. The variable $x$ is defined by $x=\sqrt{\frac{mv^2}{2\Delta E}}$, where $\frac{1}{2}mv^2$ is the kinetic energy of the impact electron and $\Delta E$ is the transition energy. For positive ions, $\bar{g}$ has a constant value of 0.2 when $x \le 1$ and is the same as the $\bar{g}$ for neutral atoms when $x > 1$. We also show the second order and the third order polynomial fit to the tabulated data. Although the third order gives a better fit to the data, it drops rapidly at larger values of $x$. Thus, we adopt the second order polynomial fit in our calculations. At high energies, the Bethe approximation shows that the collisional  excitation cross-section, like the ionization cross-section, scales as $\log E/E$.

Note that the Bethe approximation is only for permitted transitions (electric dipole transitions among states with the same spin). However, impact excitation by non-thermal electrons can occur for all transitions. To make allowance for those forbidden transitions, we use the collision strengths to compute the excitation cross sections, such that

\begin{equation}
Q_{ij} = \frac{1}{k_i^2} \frac{1}{w_i} \Omega_{ij} \pi a_0^2
\end{equation}

\noindent where $w_i$ is the statistical weight of level $i$ and $\Omega_{ij}$ is the collision strength. The main difficulty with this approach is that very few collision strengths are available. What we usually have are the thermally averaged effective collision strengths, defined by:

\begin{equation}
\Upsilon_{ij} = \int_0^{\infty} \Omega_{ij} \exp{\left(-\frac{E_j}{kT}\right)} d\left(\frac{E_j}{kT} \right)
\end{equation}

\noindent where $E_j$ is the kinetic energy of the outgoing electron. Fortunately, the thermally averaged effective collision strengths for forbidden transitions are almost constant at high temperatures, and hence we adopt such constant effective collision strengths as the collision strengths for impacting non-thermal electrons at all energies. This approximation is unlikely to have a major influence on the results since at high energies collisions for permitted transitions dominate (see Section \ref{sec:cmp_ion}), while at low energies there are relatively few non-thermal electrons, and collision rates are dominated by thermal electrons. An important distinction from non-thermal ionization is that non-thermal excitation does not produce secondary electrons.

The accuracies and the uncertainties of the Arnaud $\&$ Rothenflug cross sections and the Bethe approximation will be discussed in Section \ref{sec:exc_ion_crosec}.

\section{The hydrodynamical input}
\label{sec:hydro}

The hydrodynamical model `lm18a7Ad' (Woosley, private communication) we use as an initial input is the same as that adopted by \citet{2010MNRAS.405.2141D}. We briefly summarize the main properties of this model below. The hydrodynamical model was produced by the code {\sc kepler} \citep{1978ApJ...225.1021W}, using a progenitor with main-sequence mass of 18 M$_{\odot}$. The metallicity of the Large Magellanic Cloud (LMC) was adopted, which is $Z = 0.4 Z_{\odot}$. The star is a BSG when the explosion is initiated.

The hydrodynamical model at 0.27 day was employed, since at this time homology is a good approximation in the outer regions. The rest of the ejecta is enforced to be strictly homologous to accommodate {\sc cmfgen}, although the innermost part of the ejecta is trimmed since it suffers fallback. Moderate mixing is induced manually to soften the composition gradients. Hydrogen is deficient below 1500 \kms, while $^{56}\textrm{Ni}$ is primarily found below 2000 \kms. The total amount of $^{56}\textrm{Ni}$ in the initial hydrodynamical model is 0.084\,M$_{\odot}$. Although a large set of model atoms are utilized, some potentially important species such as Ti\,{\sc i}, Fe\,{\sc i}, Sc\,{\sc i} and Sc\,{\sc ii} are missing. The pre-SN steady and explosive nucleosynthesis in the {\sc kepler} model were undertaken with a 13-isotope network which only captures the composition approximately. It also only computes explicitly the abundance of one unstable isotope ($^{56}\textrm{Ni}$) -- numerous intermediate-mass elements (IMEs) and iron-group elements (IGEs) are bundled together. Moreover, the hydrodynamical model is enforced to be homologous, which affects the kinematics of the inner ejecta. With the recent availability of a model atom for Fe\,{\sc i} we discuss the influence of Fe\,{\sc i} on the spectra in Section~\ref{sec:fei}.

The models from day 0.27 to day 20.8 have been presented in \citet{2010MNRAS.405.2141D}, focusing on early evolution at the photospheric phase.

\section{The non-thermal model}
\label{sec:ntmodel}

To study the long-term evolution of our SN model, we evolved the model further in time until day $\sim$ 1000 -- well into the nebular phase. After about 40.6 days, the sequence is separated into two branches, a ``thermal sequence'' with $\gamma$-ray deposition as heat exclusively, and a ``non-thermal sequence'' with non-thermal excitation and ionization.

In this section, we focus on a typical non-thermal model and discuss the properties of non-thermal processes in the model. We select a non-thermal model at day 127 (hereafter model D127\_NT) near the beginning of the nebular phase. It is during the nebular phase that non-thermal effects will be the most important. Moreover, complications arising from dust will not be present since dust formation is not thought to happen until day 300 \citep{1990AJ.....99..650S,1990PNAS...87.4354G}.

\begin{figure}
  \centering
  \includegraphics[bb=35 215 585 565, width=0.45\textwidth]
  {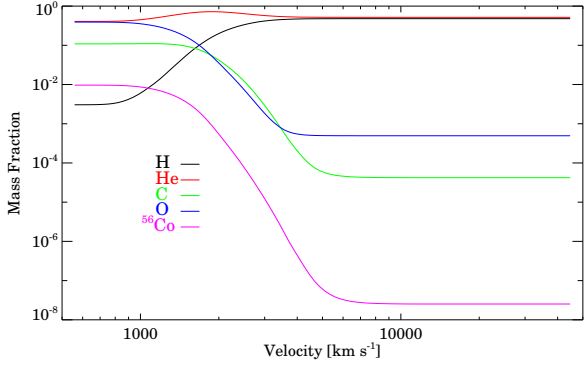}
  \caption{Illustration of the elemental mass fractions on a logarithmic scale as a function
  of velocity at day 127. We plot hydrogen (black), helium (red), carbon (green), oxygen
  (blue) and $^{56}$Co (magenta) -- five important elements in the model.}
  \label{fig:elem_pop}
\end{figure}

In Fig.~\ref{fig:elem_pop} we illustrate the elemental mass fractions on a logarithmic scale as a function of velocity at day 127. Hydrogen is an order of magnitude less abundant at velocity 1500 \kms\ and this velocity can be taken to indicate the velocity down to which H has been mixed. Rather than $^{56}\textrm{Ni}$, we show the mass fraction of $^{56}\textrm{Co}$, since at day 127 only a small amount of $^{56}\textrm{Ni}$ is left, and $^{56}\textrm{Co}$ decay is the main energy source. The figure shows that $^{56}\textrm{Co}$ has been mixed out to about 2000 \kms\ in the initial model, although trace amounts also exist at higher velocities.

\subsection{The degradation spectrum}\label{sec:deg_spec}

\begin{figure}
  \begin{center}
  \leavevmode
  \includegraphics[bb=40 230 595 550, width=0.45\textwidth]
  {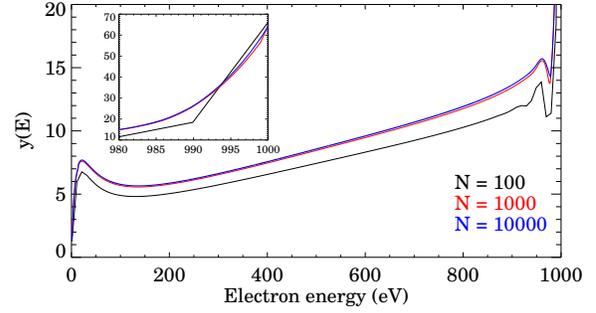}
  \caption{The degradation spectra at a model velocity of $\sim$ 1000 \kms\ computed using different numbers of energy bins: 100 (black), 1000 (red), 10\,000 (blue).}
  \label{fig:deg_spec}
  \end{center}
\end{figure}

A typical degradation spectrum at velocity $\sim$ 1000 \kms\ for the D127\_NT model is shown in Fig.~\ref{fig:deg_spec} (red).  The sharp rise at high energy is a result of the adopted source function -- we inject all fast electrons at E$_{\textrm{max}}$. The spectrum from E$_{\textrm{max}}$ = 1\,keV down to 100\,eV shows the slowing down of primary electrons, while the rise at low energy indicates a cumulation of secondary electrons. The shape of the degradation spectrum, particularly at low energies, is influenced by the ionization state of the gas.

Uncertainties in the degradation spectrum are introduced by the choice in the  number of energy bins, accuracies of the excitation and ionization cross sections, and by the choice of the high energy cut, E$_{\textrm{max}}$. The latter two will be discussed in Section \ref{sec:discussion}. For the number of energy bins, we typically adopt N = 1000 extending from 1\,eV to 1\,keV. In Fig.~\ref{fig:deg_spec}, we also show another two degradation spectra computed with energy bins N = 100 (black) and N = 10\,000 (blue). The flux in the N = 100 degradation spectrum is systematically lower than the other two, except at the high energy edge. The N = 10\,000 spectrum is very close to the one with N = 1000, and hence the uncertainty arising from using 1000 energy bins is negligible. We also made spectral comparisons among the above three models. The models with 1000 and 10\,000 energy bins showed no distinguishable differences, while the model with 100 energy bins was only slightly different from the other two. Interestingly, we later discovered that a smaller number of bins can be used, provided we scale the degradation spectrum so that the fractional energies entering the three channels sum to unity.

\subsection{Number density of non-thermal electrons}

Knowing the degradation spectrum $y(E)$, the electron spectral density in space can be estimated from $y(E)/v$, where $v = \sqrt{2E/m_{e}}$. The number density of the non-thermal electrons is then given by

\begin{equation}
N_{e} = \int_{0}^{E_{\textrm{max}}} \frac{y(E)}{v} dE
= \int_{0}^{E_{\textrm{max}}} y(E)\sqrt{\frac{m_{e}}{2E}} dE
\end{equation}

\begin{figure}
  \centering
  \includegraphics[bb=30 240 590 545, width=0.45\textwidth]
  {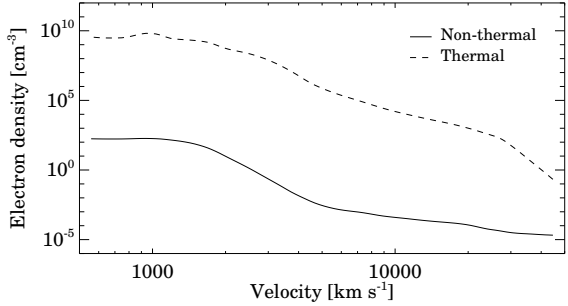}
  \caption{Comparison of the number density of the non-thermal electron (solid)
  and the thermal electron (dashed).}
  \label{fig:elec_den}
\end{figure}

Fig.~\ref{fig:elec_den} shows the comparison of number density of the non-thermal and thermal electrons  at day 127. The number density of the non-thermal electrons is many orders of magnitude smaller than that of the thermal electrons at all depths. The increase of the non-thermal electrons number density at inner regions is consistent with the increase of the $^{56}$Co abundance shown in Fig.~\ref{fig:elem_pop}.

\begin{figure}
  \centering
  \includegraphics[bb=40 240 585 545, width=0.45\textwidth]
  {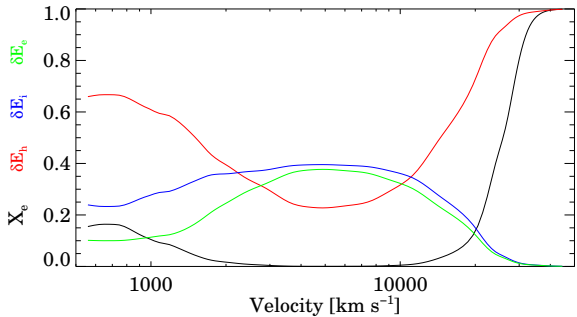}
  \caption{The fractional energy that goes into heating $\delta E_h$(red), ionization
  $\delta E_i$ (blue) and excitation $\delta E_e$ (green) in model D127\_NT,
  as a function of velocity. The ejecta ionization fraction X$_e$, which is defined as the ratio
  of number of ions to number of atoms, is also shown (black).}
  \label{fig:DMN65_heatfrac_vs_vel}
\end{figure}

\subsection{Energy fraction of the three channels}

With the inclusion of non-thermal excitation and ionization, $\gamma$-rays deposit energy into all three channels. In Fig.~\ref{fig:DMN65_heatfrac_vs_vel}, we plot the fraction of energy entering the three channels as a function of velocity. We also plot the ejecta ionization fraction X$_e$, which shows that the heating fraction is sensitive to X$_e$. In the outermost region, the heating fraction reaches unity as X$_e$ becomes 1 -- the highest ionization stage ions are unable to be ionized and excited, quenching the excitation and ionization processes. The high ionization fraction in this region is set by ionization freeze-out due to the time-dependent effects \citep{2008MNRAS.383...57D}. Between 3000 \kms\ and 10\,000 \kms, non-thermal excitation and ionization become prominent where X$_e$ becomes very small. As X$_e$ increases inward from 3000 \kms, fractional non-thermal excitation and ionization decline and fractional heating rises and becomes dominant.

\subsection{Excitation and ionization}\label{sec:exc_ion_cool}

When non-thermal excitation and ionization occur, energy from non-thermal electrons is used to excite or ionize an atom. The Spencer-Fano equation (Eq.~\ref{eq:spencer}) implies that the abundances and the cross sections are the key to determine what energy a species ``consumes''. In this subsection, we study the fractional energies consumed by a species and compare them to the corresponding species.

\begin{figure}
  \subfigure{\includegraphics[bb=40 220 580 560, width=0.48\textwidth]
  {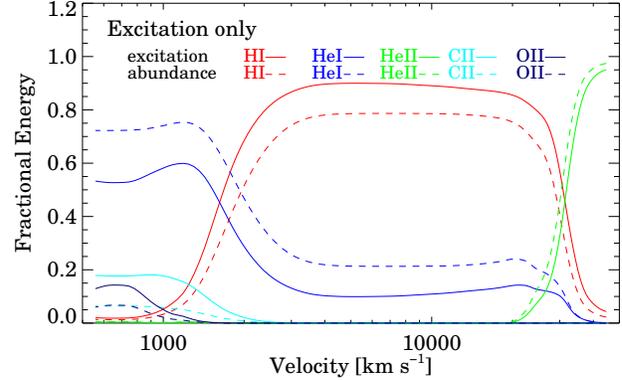}}
  \subfigure{\includegraphics[bb=40 220 580 560, width=0.48\textwidth]
  {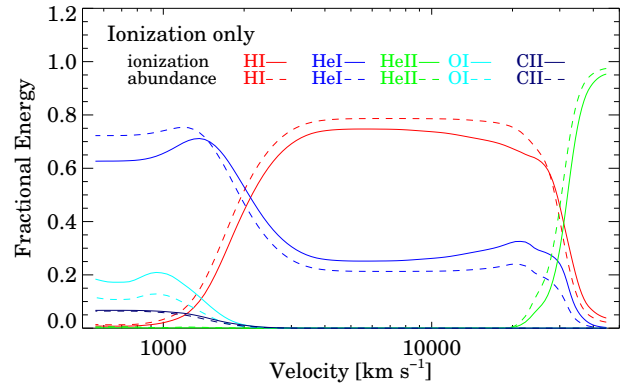}}
  \caption{The fractional energy (solid lines) consumed by a species in the
  non-thermal excitation (top) and ionization (bottom) processes as a function
  of ejecta velocity. Only the largest 5 (summation over all depths) fractions
  are shown. The fractions of species abundances are plotted in dashed lines for
  comparison. Importantly, the highest ionization stage of each element is
  excluded in the calculations of fractional abundances. (see the text for
  details).}
  \label{fig:popfrac_info}
\end{figure}

In Fig.~\ref{fig:popfrac_info}, we illustrate the fractional energies taken away from the non-thermal electrons by a species in both the excitation and ionization channels. The largest 5 consumers, summing over all depths, are plotted (solid lines). We overplot the fractional abundances of these 5 species (dashed lines). However, the calculations of fractional abundances excludes the highest ionization stages of all elements. These highest ionization stages are the ones in the model, not necessarily the same as the ion of an element with no electron left. For instance, in our models,  we utilize detailed model atoms for  Fe\,{\sc ii}, Fe\,{\sc iii}, Fe\,{\sc iv}, Fe\,{\sc v}, Fe\,{\sc vi} and Fe\,{\sc vii}, then Fe\,{\sc viii} is the highest ionization stage for iron. The reason for excluding the highest ionization stages is that we treat these ionization stages as a single level, and so they don't have any excitation or ionization routes. In both channels, it is obvious that H\,{\sc i}, He\,{\sc i} and He\,{\sc ii} are the dominant consumers of non-thermal electron energy. He\,{\sc ii} is the main consumer above $30\,000$~\kms. In the middle region, both H\,{\sc i} and He\,{\sc i} are prominent.  Due to low abundance of H, He\,{\sc i} dominates the inner zone, although other species, like C\,{\sc i}, C\,{\sc ii}, O\,{\sc i} and O\,{\sc ii} also play a role. Fig.~\ref{fig:popfrac_info} shows that the species abundances, especially H\,{\sc i}, He\,{\sc i} and He\,{\sc ii}, have similar patterns to their fractional energies, in both channels. Such similarity indicates that allocation of the non-thermal energy is generally controlled, as expected, by the species abundances.

\section{Non-thermal vs thermal models}\label{sec:mod_cmp}

In this section, we present a comparison of the temperature and ionization structure between the non-thermal and thermal models. The spectral evolution of the two sequences will also be compared.

\subsection{Comparison of the temperature structure}\label{sec:cmp_temp}

\begin{figure}
  \begin{center}
  \leavevmode
  \includegraphics[bb=40 255 585 530, width=0.45\textwidth]
  {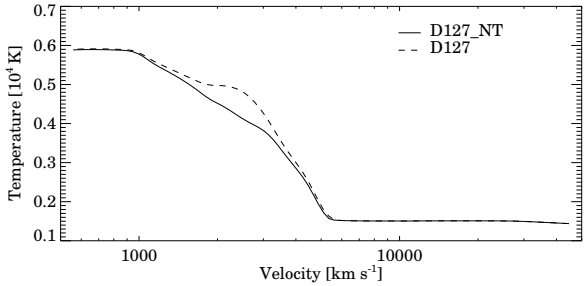}
  \caption{Comparison of temperature structure between model D127\_NT
  (solid) and model D127 (dashed).}
  \label{fig:temp_vs_vel}
  \end{center}
\end{figure}

In Fig.~\ref{fig:temp_vs_vel}, we compare the temperature structure between model D127\_NT and the thermal model at day 127 (hereafter model D127). As mentioned previously, the non-thermal model generally has a lower temperature in the region between 1000 \kms\ and 5000 \kms\ where the non-thermal effects are crucial. The temperature for the two models above 5000 \kms\ is pretty flat (i.e., almost constant). This is an artifact -- a floor temperature of 1500 K is imposed for the models to prevent numerical overflow. The flux mean opacity at 5000 \kms\ is of order $10^{-2}$, thus the region above 5000 \kms\ has very little influence on spectral formation\footnote{This is confirmed by computing two models with different floor temperatures. The synthetic spectra show no distinguishable difference between models with floor temperatures of 1500 K and 4000 K.}.

The decrease in the temperature can be understood as follows. The thermal model puts all decay energy into heating, while the non-thermal model only puts a portion of decay energy into heating. Moreover, the coupling between the gas and the radiation at this time is relatively weak, and hence, we expect that only the heating channel will substantially affect the temperature. Energy that has not gone into heat/temperature is stored up in ionization/excitation, driving the material even further away from LTE level populations.

Notice that the temperature of the two models is similar below 1000 \kms, despite the existence of a substantial amount of $^{56}\textrm{Ni}$. Even though the region is not in LTE, collisional processes are important, and there is a strong coupling between the radiation field and the gas, and thus energy initially deposited via the ionization and excitation channels can be thermalized. The fractions of energy entering the three channels are shown in Fig.~\ref{fig:DMN65_heatfrac_vs_vel}. The heating fraction becomes dominant in the inner region, which is consistent with the temperature comparison in Fig.~\ref{fig:temp_vs_vel}.

\subsection{Controlling processes for H\,{\sc i} and He\,{\sc i} lines}\label{sec:cmp_ion}

\begin{figure}
  \subfigure{\includegraphics[bb=40 220 585 565, width=0.45\textwidth]
  {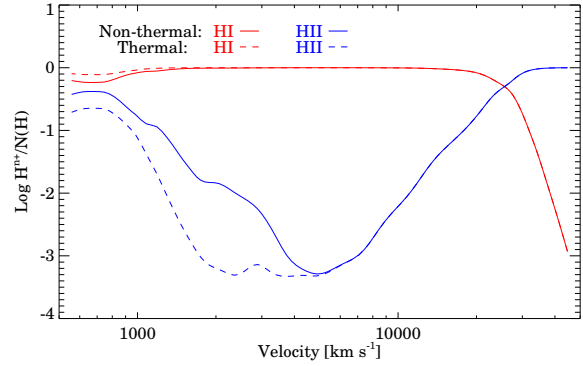}}
  \subfigure{\includegraphics[bb=40 220 585 565, width=0.45\textwidth]
  {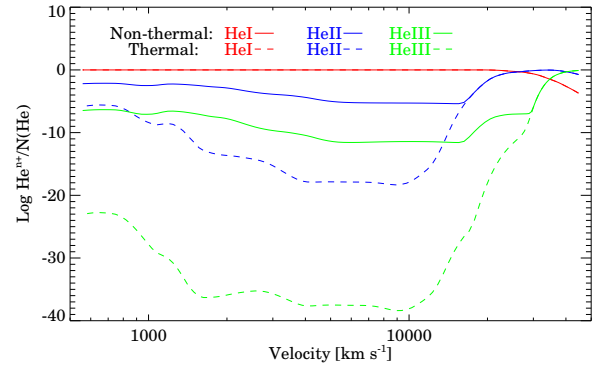}}
  \caption{Logarithmic species fractions of different ionization stages for
  hydrogen (top) and helium (bottom). Comparison is made between model D127\_NT
  (solid lines) and D127 (dashed lines).}
  \label{fig:cmp_specfrac}
\end{figure}

In this section, we compare the hydrogen and helium ionization structure in the D127 and D127\_NT models and study the  processes controlling the H\,{\sc i} and He\,{\sc i} lines in the non-thermal model. The H and He ionization fractions for both the D127 and D127\_NT models are illustrated in Fig.~\ref{fig:cmp_specfrac}. In the cobalt region, the ionized H fraction has increased by up to an order of magnitude, while in the region above $v > 5000$ \kms, the ionization is unchanged, since $^{56}\textrm{Co}$ is deficient there.

While He remains neutral for $v < 10\,000$~\kms, the singly ionized He fraction is now significant. We can also see significant change in the He ionization fraction above the region where cobalt is abundant, which is due to a small amount of cobalt in the model. At large velocities, the ionization of both H and He is frozen, since the expansion time-scale is
smaller than the recombination time-scale \citep{2010MNRAS.405.2141D}.

\begin{figure}
  \begin{center}
  \leavevmode
  \includegraphics[bb=0 160 595 635, width=0.48\textwidth]
  {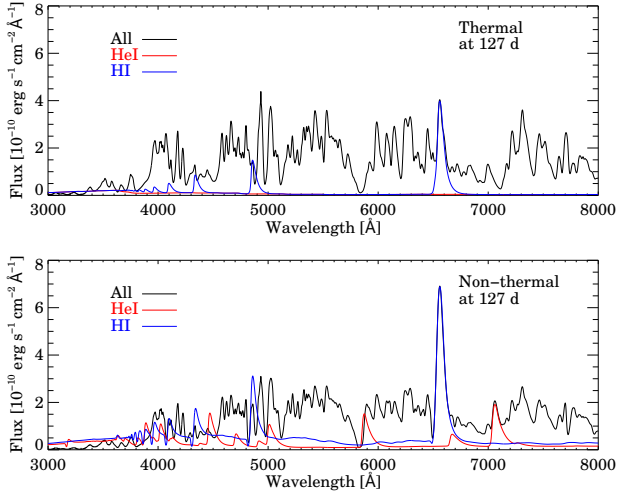}
  \caption{Upper panel: The D127 model spectrum (black) and the model spectra
  computed by including only bound-bound transitions of He\,{\sc i} (red) and
  H\,{\sc i} (blue) at optical band. Lower panel: Same as the upper panel, but
  for the D127\_NT model.}
  \label{fig:HeI_spectra}
  \end{center}
\end{figure}

\begin{figure}
  \begin{center}
  \leavevmode
  \includegraphics[bb=15 160 595 640, width=0.48\textwidth]
  {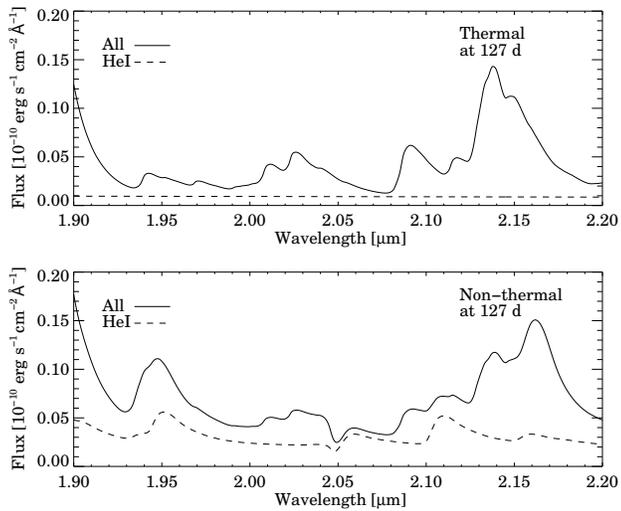}
  \caption{Illustration of model spectrum (solid) and model spectrum computed
  by including only bound-bound transitions of He\,{\sc i} (dashed) for model D127
  (top panel) and D127\_NT (bottom panel) between 1.9 and 2.2\,$\mu$m. }
  \label{fig:HeI_spectra_HeI20580}
  \end{center}
\end{figure}

Fig.~\ref{fig:HeI_spectra} shows the model spectra and the spectra with only bound-bound transitions of He\,{\sc i} and H\,{\sc i} at the optical band. Optical, as well as IR, H\,{\sc i} and He\,{\sc i} lines mainly originate below 2000 \kms\ in the non-thermal model, with a small fraction of the line flux coming from 2000 - 3000~\kms\ for some lines, e.g., He\,{\sc i} 1.083\,$\mu$m. H\,{\sc i} lines in the thermal model (upper panel) are generally weak and He\,{\sc i} lines are absent. In the non-thermal model (lower panel), all H\,{\sc i} lines are strengthened, with H$\alpha$ particularly boosted by non-thermal processes. He\,{\sc i} lines also contribute significantly to the spectrum. While most optical He\,{\sc i} lines are contaminated by other lines, He\,{\sc i}\,7065\,\AA\ might be a good diagnostic to infer the influence of non-thermal processes on helium. Notice that He\,{\sc i} 4471\,\AA\ is excited in the non-thermal model but is absent from the spectra due to line-blanketing.

\begin{figure}
  \subfigure{\includegraphics[bb=110 260 505 500, width=0.48\textwidth]
  {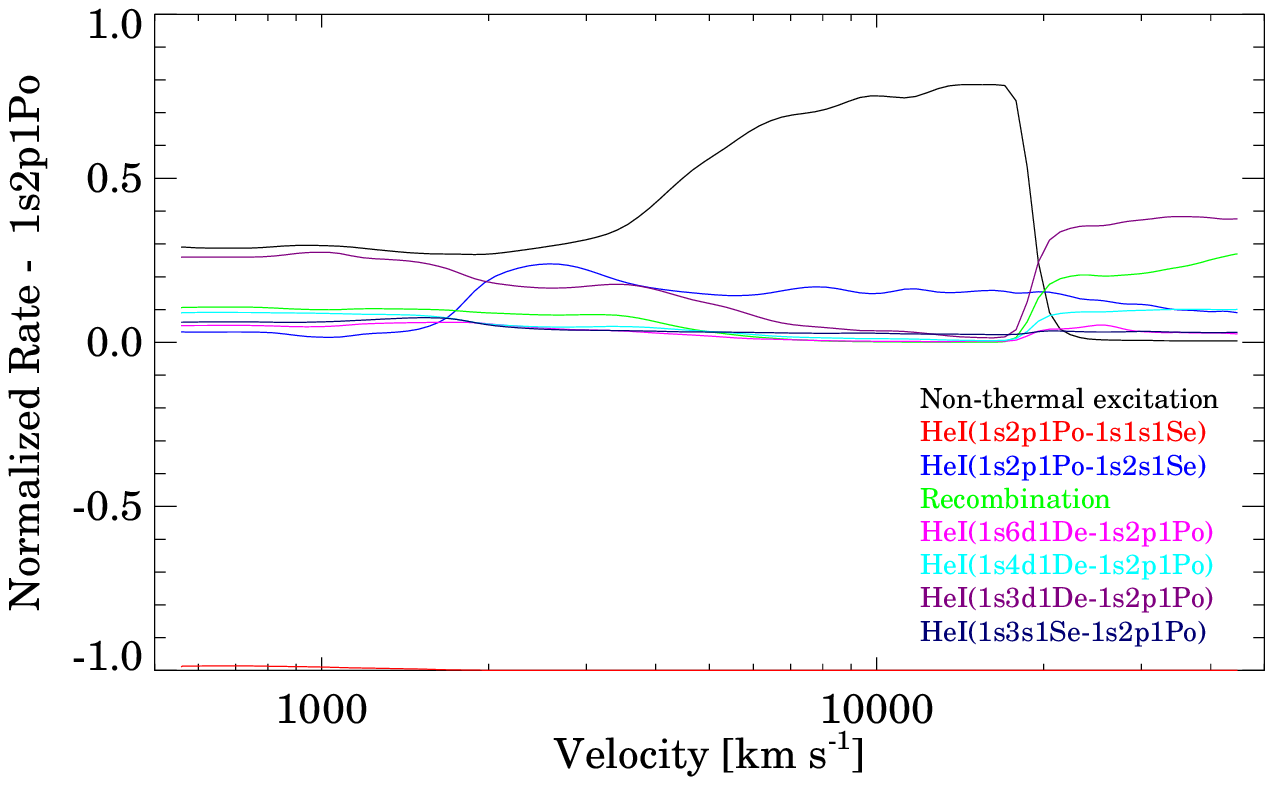}}
  \subfigure{\includegraphics[bb=110 260 505 500, width=0.48\textwidth]
  {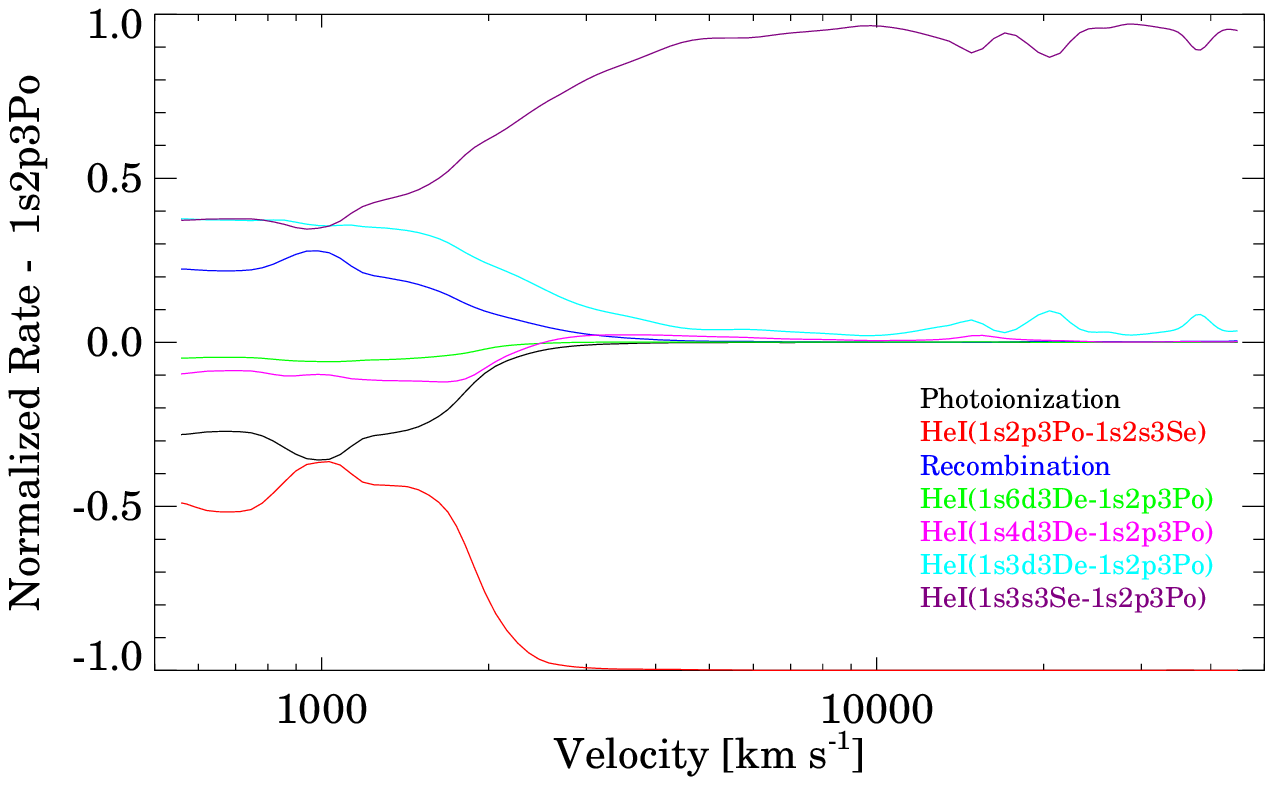}}
  \caption{Top: Normalized rates for different processes that populate and depopulate the
  1s\,2p\,$^1$P$^{\scriptsize o}$ state of He\,{\sc i}. Only processes with a fractional rate
  $\ge$ 5\% at more than one depth are shown. Bottom: Same as the top panel, but for
  the case of 1s\,2p\,$^3$P$^{\scriptsize o}$ state of He\,{\sc i}.}
  \label{fig:mod_rate}
\end{figure}

The prominence of He\,{\sc i}\,1.083\,$\mu$m and He\,{\sc i}\,2.058\,$\mu$m due to non-thermal processes at late times (t $\ge$ 200 d) was noted by \citet{1995ApJ...441..821L}, and observations showed He\,{\sc i}\,2.058\,$\mu$m is more isolated \citep{1989MNRAS.238..193M}. In Fig.~\ref{fig:HeI_spectra_HeI20580}, we show the model spectra and the spectra with only bound-bound transitions of He\,{\sc i} for models D127\_NT and D127, focusing on the He\,{\sc i}\,2.058\,$\mu$m line. Similar to the optical band, no He\,{\sc i} line is present in model D127 in the wavelength range from 1.9 $\mu$m to 2.2 $\mu$m. However, He\,{\sc i}\,2.058\,$\mu$m produces a prominent absorption feature in the D127\_NT model. This feature has been seen in observations of SN 1987A \citep{1989MNRAS.238..193M}, providing strong evidence for the influence of non-thermal processes on He\,{\sc i}. \citet{1995ApJ...441..821L} showed that He\,{\sc i}\,2.058\,$\mu$m  is mainly due to non-thermal excitation and He\,{\sc i}\,1.083\,$\mu$m is mainly excited by thermal electrons during 200\,d $\le$ t $\le$ 450\,d.

\begin{figure}
  \begin{center}
  \subfigure{\includegraphics[bb=100 270 520 520, width=0.48\textwidth]
  {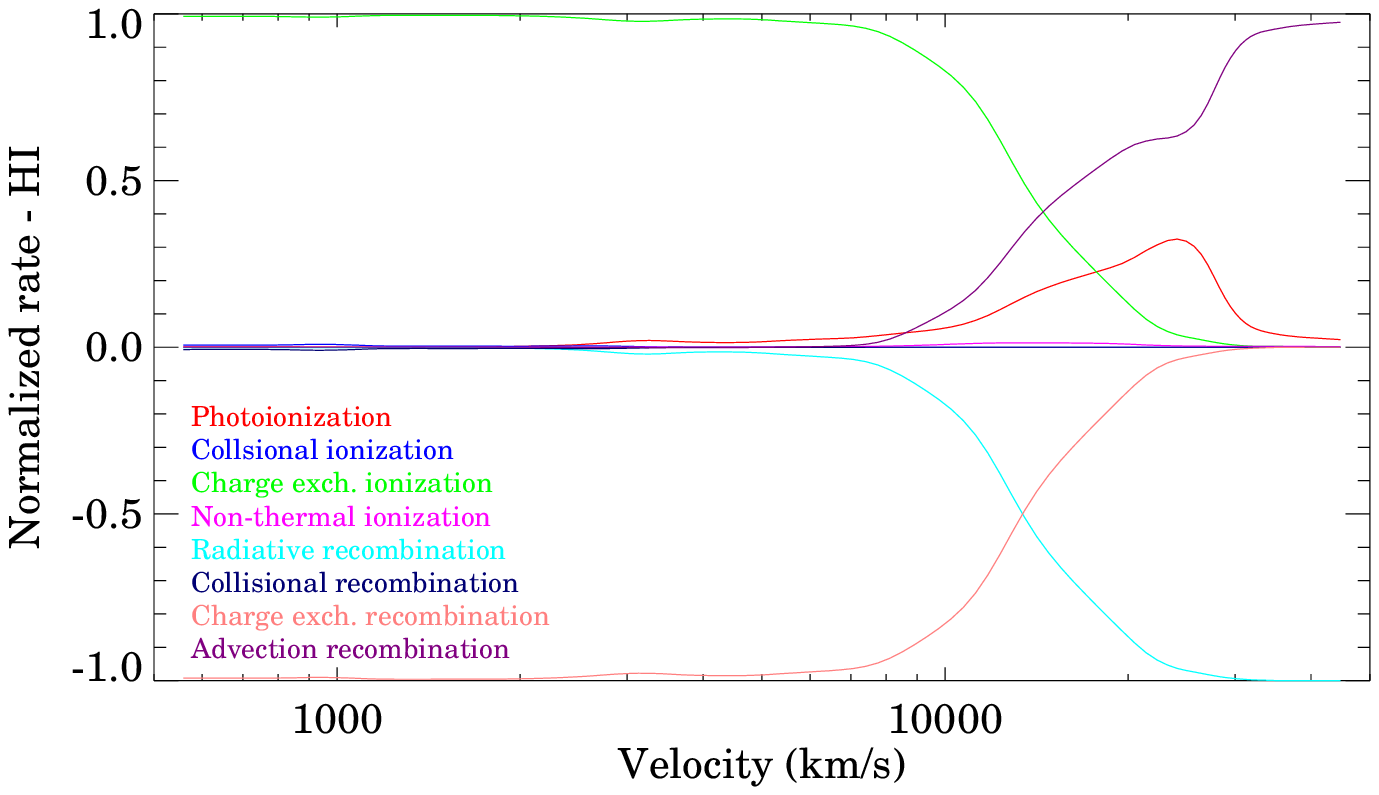}}
  \subfigure{\includegraphics[bb=100 270 520 520, width=0.48\textwidth]
  {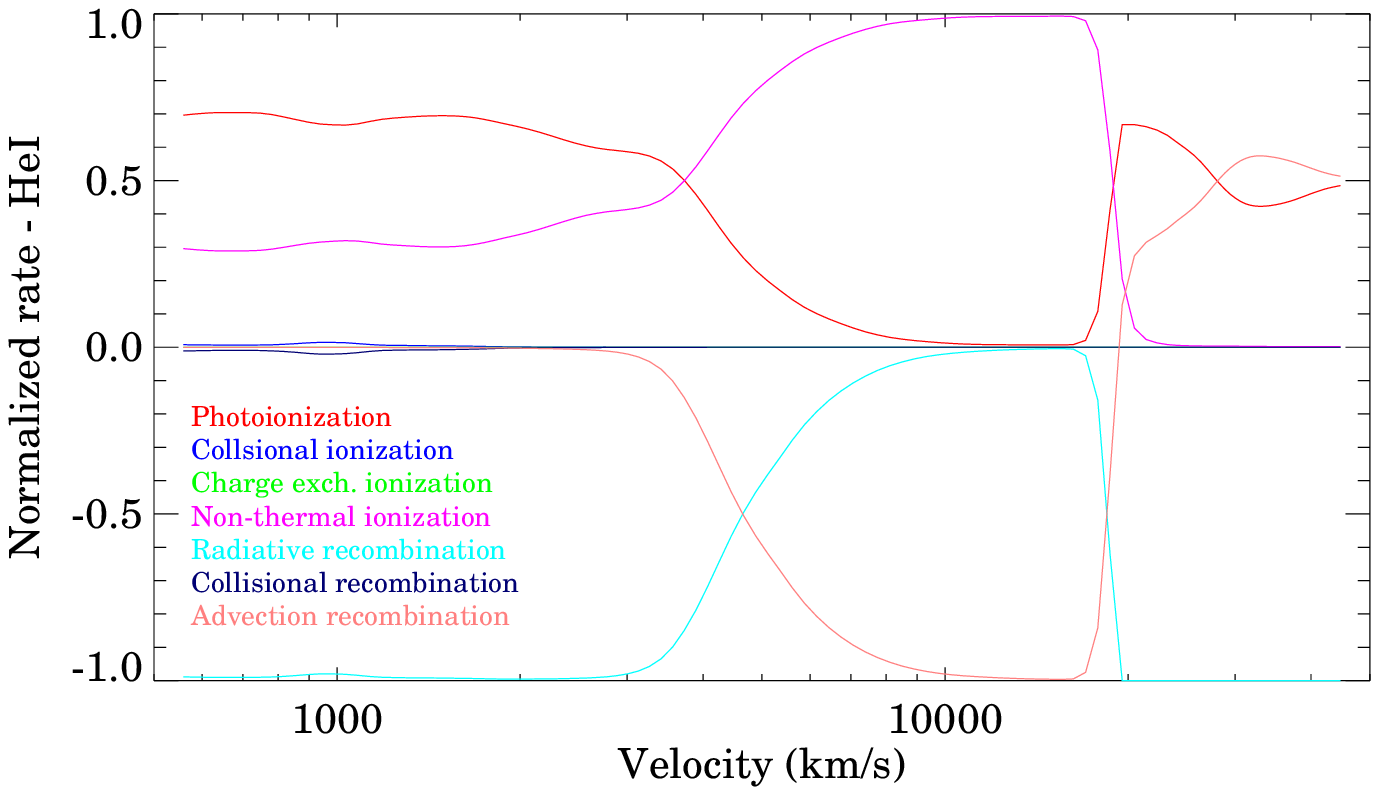}}
  \caption{Normalized rates for H\,{\sc i} (top) and He\,{\sc i} (bottom), including all
  processes allowed in the modeling.}
  \label{fig:mod_prrr}
  \end{center}
\end{figure}

In Fig.~\ref{fig:mod_rate}, we illustrate the major fractional rates ($\ge$ 5\% at any depth) that populate and depopulate the states 1s\,2p\,$^1$P$^{\scriptsize o}$ and 1s\,2p\,$^3$P$^{\scriptsize o}$ of He\,{\sc i} for model D127\_NT. Positive normalized rates illustrate the routes that populate the state, while negative ones show the routes that depopulate the state. States 1s\,2p\,$^1$P$^{\scriptsize o}$ and 1s\,2p\,$^3$P$^{\scriptsize o}$ are the upper levels for the He\,{\sc i} lines 2.058\,$\mu$m and 1.083\,$\mu$m, respectively. For the state 1s\,2p\,$^1$P$^{\scriptsize o}$, we confirm that non-thermal excitation is the main excitation mechanism. Cascades from higher levels, especially the 1s\,3d\,$^1$D state, also play an important role at this stage.

For the state 1s\,2p\,$^3$P$^{\scriptsize o}$, cascades from higher levels (mainly 1s\,3s\,$^3$S and 1s\,3d\,$^3$D) are the key processes to populate it below 3000 \kms. Recombination is also non-negligible. These are indirectly related to the non-thermal ionization processes. Thus, in contrast to the He\,{\sc i}\,2.058\,$\mu$m line, non-thermal excitation is largely irrelevant for He\,{\sc i}\,1.083\,$\mu$m. This is due to the difference in the collision strength behavior of the two transitions. Non-thermal excitation from the ground state (1s\,1s\,$^1$S) is generally the dominant excitation route, since the ground state is the most populated state. At low energies, the collision strength for the singlet transition (permitted transition), $\Omega$(1s\,1s\,$^1$S\,-\,1s\,2p\,$^1$P$^{\scriptsize o}$), is similar to that of the triplet transition (forbidden transition), $\Omega$(1s\,1s\,$^1$S\,-\,1s\,2p\,$^3$P$^{\scriptsize o}$). However, $\Omega$(1s\,1s\,$^1$S\,-\,1s\,2p\,$^1$P$^{\scriptsize o}$) grows significantly as the impact energy increases, while $\Omega$(1s\,1s\,$^1$S\,-\,1s\,2p\,$^3$P$^{\scriptsize o}$) is approximately constant at high energies. As a result, non-thermal excitation processes contribute much less to the triplet states than to the singlet states. Due to radiative excitation from 1s\,2s\,$^3$S to 1s\,3p\,$^3$P$^{\scriptsize o}$ and subsequent decay to 1s\,3s\,$^3$S, electron cascade from 1s\,3s\,$^3$S, rather than 1s\,3d\,$^3$D, is the dominant process  populating the 1s\,2p\,$^3$P$^{\scriptsize o}$ state above 3000 \kms.

\begin{figure}
  \subfigure{\label{fig:cmp_snspec_nt_vs_t_1}
  \includegraphics[bb=50 275 595 515, width=0.48\textwidth]
  {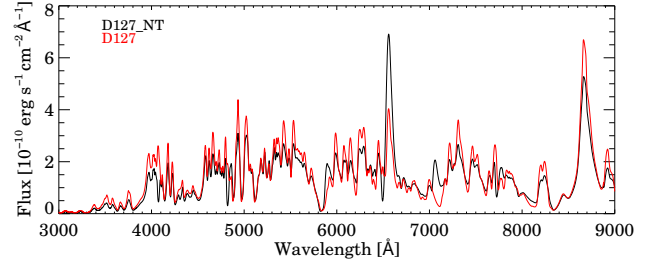}}
  \subfigure{\label{fig:cmp_snspec_nt_vs_t_2}
  \includegraphics[bb=50 275 595 515, width=0.48\textwidth]
  {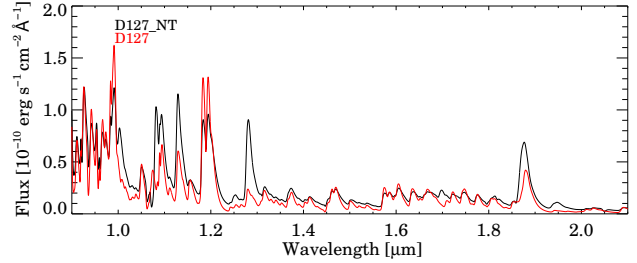}}
  \caption{Comparison of optical and IR spectra between models D127\_NT and D127.}
  \label{fig:cmp_snspec_nt_vs_t}
\end{figure}

\begin{figure}
  \begin{center}
  \leavevmode
  \includegraphics[bb=50 270 595 520, width=0.48\textwidth]
  {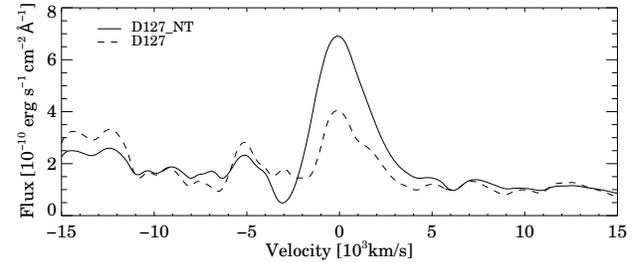}
  \caption{Comparison of the H$\alpha$ profile between models
  D127\_NT (solid) and D127 (dashed).}
  \label{fig:cmp_snspec_ha}
  \end{center}
\end{figure}

\begin{figure*}
  \begin{center}
  \includegraphics[bb=30 215 590 575, width=0.98\textwidth]
  {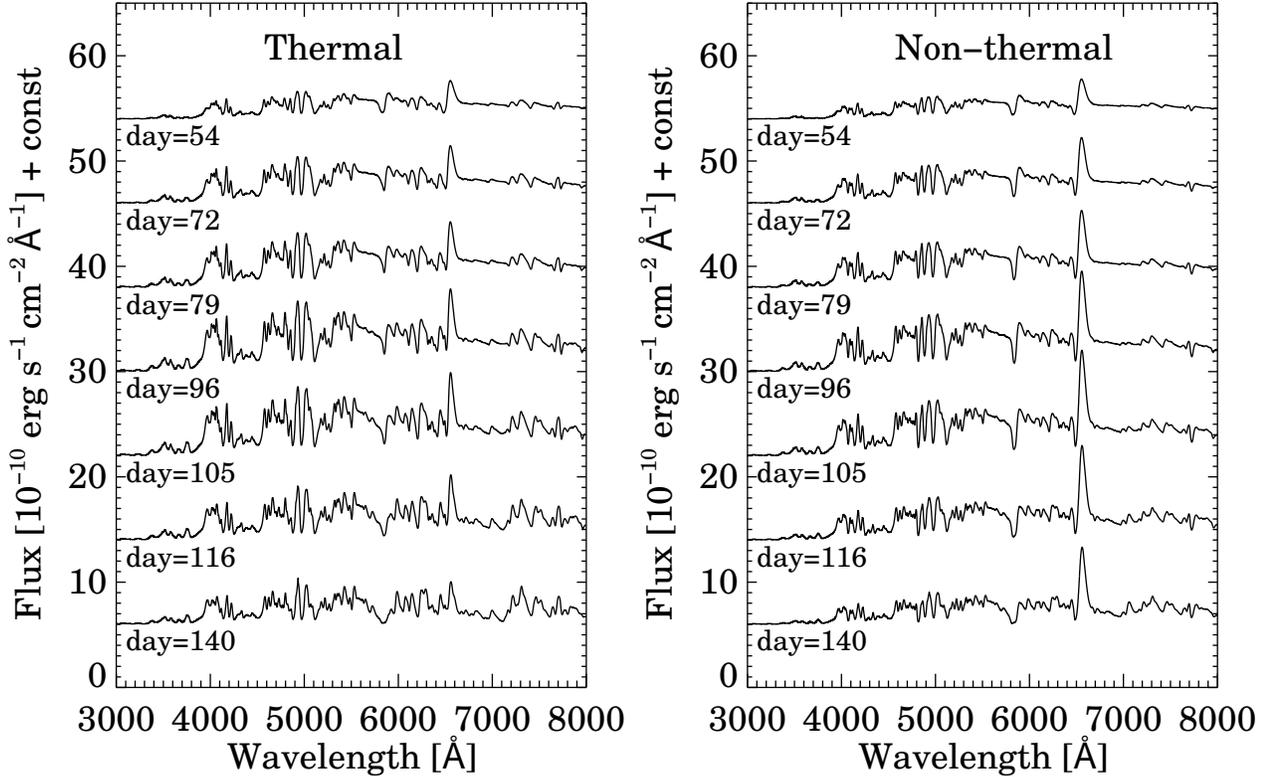}
  \caption{Left panel: montage of synthetic optical spectra for the thermal sequence.
  The days since breakout are labeled below each spectra. Right panel: montage of
  synthetic spectra at the same epochs shown in the left panel for the non-thermal
  sequence. The synthetic spectra are reddened with E(B-V) = 0.15 and scaled for
  a distance of 50 kpc. The most prominent difference between the spectra of the two
  sequences is the evolution of H$\alpha $ -- its strength persists at all times in
  the non-thermal sequence but it almost disappears at late times in the thermal sequence.}
  \label{fig:cmp_nt_evol}
  \end{center}
\end{figure*}

Despite the importance of non-thermal processes, we find that photoionization and recombination play a crucial, if not dominant,  role in populating many other levels. However, in a ``thermal'' model, the photoionization and recombination rates are too small to produce a prominent effect. The trigger for the  large photoionization and recombination rates is non-thermal processes. In Fig.~\ref{fig:mod_prrr} we show the normalized rates affecting the H\,{\sc i} and He\,{\sc i} ionization balance for model D127\_NT. For H\,{\sc i}, the fraction of non-thermal energy deposited via the ionization channel is small and photoionization and recombination are the dominant processes and they balance each other below 8000 \kms. We check an individual level, $n = 1$, of H\,{\sc i} in both models (D127\_NT and D127), and find that the photoionization and recombination rates are increased by a factor of $\sim$ 2 when non-thermal processes are included. For the He\,{\sc i} 1s\,2p\,$^1$P$^{\scriptsize o}$ state, non-thermal ionization and excitation rates in model D127\_NT are 5 orders of magnitude greater than any rate in model D127. Below 3000 \kms, non-thermal ionization becomes significant, and the radiative recombination rate is now balanced by the photoionization and the non-thermal ionization rates. Although the net non-thermal rate is not overwhelmingly dominant, it is crucial for controlling the ionization balance.

\subsection{Comparison of optical and IR spectra}

Fig.~\ref{fig:cmp_snspec_nt_vs_t} shows the comparison of the optical and IR spectra between models D127\_NT and D127. In the optical range (Fig.~\ref{fig:cmp_snspec_nt_vs_t}, top panel), D127\_NT generally produces slightly lower fluxes but with two enhanced prominent features --- H$\alpha$ and He\,{\sc i}\,7065\AA. The Ca\,{\sc ii} IR triplet is slightly weakened. In the bottom panel of Fig.~\ref{fig:cmp_snspec_nt_vs_t}, most of the lines are strengthened, such as He\,{\sc i}\,1.083\,$\mu$m, Pa $\gamma$ 1.094 $\mu$m, O\,{\sc i}\,1.129\,$\mu$m, Pa $\beta$ 1.281 $\mu$m and Pa $\alpha$ 1.875 $\mu$m. However, Ca\,{\sc ii}\,1.184\,$\mu$m and 1.195\,$\mu$m are weakened.

\cite{2011MNRAS.410.1739D} found that in their SNe II-P models the Balmer lines suddenly disappeared at end of the photospheric phase -- a result of the vanishing of Balmer-continuum photons. However, observations of Type II SNe show a strong H$\alpha$ profile during the nebular phase. They attributed this discrepancy to the exclusion of non-thermal excitation and ionization in the models. We present here the comparison of H$\alpha$ at day 127 in Fig.~\ref{fig:cmp_snspec_ha}. While H$\alpha$ is almost absent in model D127, it shows a strong P Cygni H$\alpha$ profile in model D127\_NT. The emission component is stronger and wider and a P Cygni absorption trough is now present. The width of H$\alpha$ is related to the outward mixing of $^{56}\textrm{Ni}$. This may be helpful to determine the $^{56}\textrm{Ni}$ mixing in Type II SNe, as hydrogen is abundant in these SNe and H$\alpha$ is one of the strongest features at nebular epochs.

\subsection{Comparison of the spectral evolution}
\label{sec:cmp_mod_spec}

As mentioned in the previous section, \citet{2011MNRAS.410.1739D} were unable to reproduce the strong H$\alpha$ at the nebular phase for red supergiant (RSG) models, which was associated with the neglect of non-thermal processes. Moreover, the mixing in their models may have been too weak. The best explanation for the persistence of H$\alpha$ at late times is non-thermal excitation and ionization processes combined with mixing of H and $^{56}\textrm{Ni}$. This has been one of the motivations for this study. In Fig.~\ref{fig:cmp_nt_evol}, we present the spectral evolution of the non-thermal and the thermal sequences. The disappearance of H$\alpha$ is also observed in the thermal model for SNe resulting from a blue supergiant (BSG) progenitor. The strength of H$\alpha$ increases till day 96, and then fades quickly. On day 140, no prominent feature is seen. The spectral evolution of the non-thermal sequence shows similarity, except for H$\alpha$, which has a stronger maximum strength at about day 96, decreases gradually after that, and maintains a stronger signature on day 140. This confirms that the persistence of strong H$\alpha$ emission during the nebular phase is closely related to non-thermal excitation and ionization. The non-thermal effects on spectra at day 54 are relatively small, confirming that the neglect of the non-thermal processes in earlier models is unlikely to have a significant influence on model spectra.

\begin{figure}
  \begin{center}
  \includegraphics[bb=50 270 595 520, width=0.48\textwidth]
  {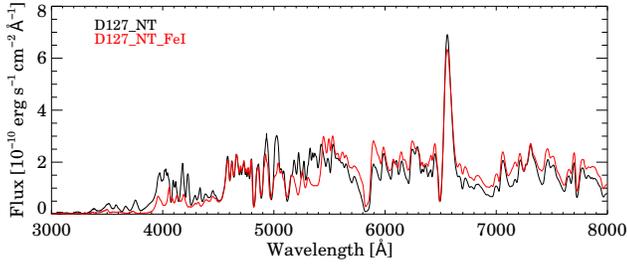}
  \caption{Comparison of optical spectral between model D127\_NT and the model
  with Fe\,{\sc i} (hereafter model D127\_NT\_FeI).}
  \label{fig:cmp_snspec_fei}
  \end{center}
\end{figure}

\section{The influence of F\lowercase{e\,{\bf \sc i}}}
\label{sec:fei}

\begin{figure}
  \begin{center}
  \includegraphics[bb=100 235 530 550, width=0.48\textwidth]
  {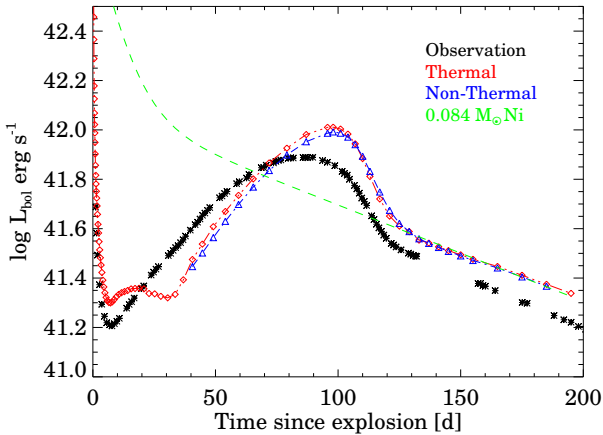}
  \caption{Comparison of the observed (black stars,
  \citeauthor{1990AJ.....99..650S} \citeyear{1990AJ.....99..650S}) bolometric light
  curve of SN 1987A to the theoretical bolometric light curves of the thermal (red)
  and non-thermal (blue) sequences, with symbols referring to the computed epochs of
  models. The observational bolometric light curve is constructed by using
  ultraviolet, optical and infrared photometry. A light curve resulting from the conversion
  of energy from the radioactive decay of 0.084 M$_{\odot}$ (green) is also shown for
  reference.}
  \label{fig:snlc_bol}
  \end{center}
\end{figure}

The source of opacity in SNe is complicated, since lots of IMEs and IGEs are synthesized during the explosion. Although a huge number of levels are included in the model, we are still missing some model atoms and the amount of levels of current model atoms may also be insufficient.
Many of the missing model ions are the lowest ionization stages. These do not influence the spectra at early times when the ejecta is hot, but may significantly influence the opacity at late times when the ejecta becomes cold. Here, we explore the influence of Fe\,{\sc i}. Oscillator strengths are from \citet{2009AIPC.1171...43K}\footnote{Data from R. L. Kurucz is available online at http://kurucz.harvard.edu}, photoionization cross sections are
from \cite{Bau97_FeI_phot} (accesed through Nahar-OSU-Radiative-Atomic-Data web site), while collision rates among the lowest 10 levels are from \cite{PB97_FeI_col} (accessed from TIPBASE at cdsweb.u-strasbg.fr/tipbase/home.html).

We included Fe\,{\sc i} at day 5 and ran a sequence of models with this new atom and with non-thermal excitation and ionization. Fig.~\ref{fig:cmp_snspec_fei} illustrates the comparison of the optical spectra at day 127. With the inclusion of Fe\,{\sc i}, the emergent flux decreases appreciably between 3000 \AA\ and 5500 \AA, with the energy  redistributed to longer wavelengths above 6000 \AA. Many other atoms, such as Co\,{\sc i}, Ti\,{\sc i} and Sc\,{\sc ii}, may cause effects similar to Fe\,{\sc i}. We will include more complete model atoms, as they become available, to better address line blanketing.

\section{Comparison with the observations}
\label{sec:obs_cmp}

\begin{figure}
  \begin{center}
  \includegraphics[bb=100 210 530 570, width=0.48\textwidth]
  {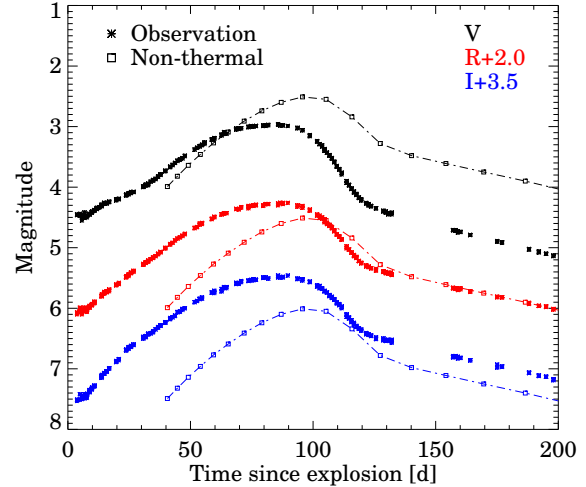}
  \caption{Comparison of the V-, R-, and I-band light curves of SN 1987A to the corresponding
  synthetic light curves. The observed V, R, and I magnitudes are shown in stars, and the
  synthetic V, R, and I magnitudes from the non-thermal sequence are plotted in squares
  connected by dashed-dotted lines. A scaling is applied to the R and I band magnitudes, both observed and synthetic, to optimize visualization. The synthetic photometry is computed by convolving the synthetic spectra with the Landolt filter bandpasses \citep{1992AJ....104..340L}. We applied a reddening of E(B-V)=0.15 and adopted a distance of 50 kpc.}
  \label{fig:snlc_ubvri}
  \end{center}
\end{figure}

SN 1987A is still one of the best observed and studied SNe, and it is about 50 kpc away from the earth. Although its progenitor was a BSG, there is still debate about the evolutionary channel that led to a SN explosion as a BSG  \citep{1987Natur.327..597H,1989A&A...224L..17L,1990A&A...227L...9P}. In this section, we present photometric and spectroscopic comparison with the observations of SN 1987A. Our observational data is taken from CTIO \citep{1988AJ.....95.1087P}. For the synthetic spectra, we adopt the extinction curve of \cite{1989ApJ...345..245C} and use a reddening of E(B-V)$=0.15$. \cite{2005coex.conf..585P} derived the distance to SN 1987A to be 51.4~kpc with an uncertainty of 1.2~kpc. We adopt $d=50$~kpc for consistency with \citet{2010MNRAS.405.2141D}.

\begin{figure*}
  \begin{center}
  \leavevmode
  \includegraphics[bb=30 300 595 480, width=0.98\textwidth]
  {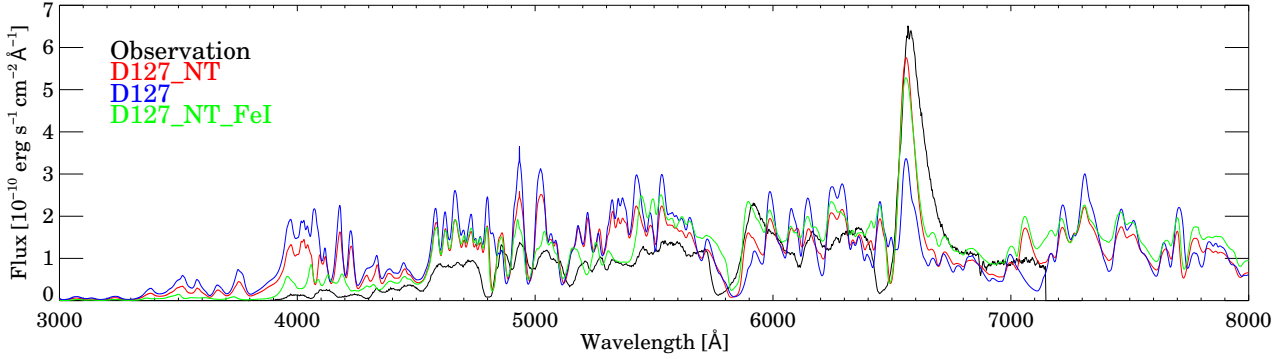}
  \caption{Comparison between observed spectrum of SN 1987A and the model spectra
  at day 127. A scaling is applied to the observed spectrum to correct for the difference
  between the spectroscopic (4.30) and the photometric (4.37) V-magnitude. The
  three model spectra are scaled by 0.07/0.084 according to the $^{56}\textrm{Ni}$
  mass in SN 1987A and in the model.}
  \label{fig:cmp_snspec_3000_8000}
  \end{center}
\end{figure*}

\subsection{The synthetic and observed light curves}
\label{sec:light_curves}

The comparison of the bolometric light curves is illustrated in Fig.~\ref{fig:snlc_bol}, while Fig.~\ref{fig:snlc_ubvri} illustrates the comparison of the V-, R-, and I-band light curves. The theoretical bolometric light curve rebrightens at a later time than observation, which is probably due to insufficient outward mixing of $^{56}\textrm{Ni}$ \citep{2004AstL...30..293U,2012inprep0000.0001D}. The delay of the predicted peak luminosity supports this idea. Both the thermal and non-thermal bolometric light curves predict a higher luminosity than the observed one, partly due to higher $^{56}\textrm{Ni}$ mass in our models. \citet{1990AJ.....99..650S} derived a $^{56}\textrm{Ni}$ mass of 0.07\,M$_{\odot}$, while we have 0.084\,M$_{\odot}$ in the model, which is the amount that comes out of the explosion model. As the purpose of this work is to explore the influence of non-thermal processes, and not to fit the spectra of SN\,1987A, we did not adjust the progenitor model, nor did we explore the influence of mixing, explosion energy, and mass cut in the progenitor model. The difference in the amount of $^{56}\textrm{Ni}$ only affects our results quantitatively, not qualitatively. The hydrodynamical model, such as the size of the helium core and the enforcement of homology at the beginning of the modeling, also introduces uncertainties. Considering that we have no free parameter, the synthetic bolometric light curves agrees reasonably well with the observed bolometric light curves. We also plot a light curve resulting from the radioactive decay of 0.084\,M$_{\odot}$ of $^{56}\textrm{Ni}$. The agreement between this light curve and the synthetic bolometric light curve during the nebular epoch demonstrates that the energy source of the ejecta is totally dominated by radioactive decay.

Although the synthetic bolometric light curve is reasonably well reproduced, the synthetic multi-band light curves show varying degrees of systematic offsets. We are mainly interested in the nebular epochs when the impact of mixing is less important. The R-band magnitudes are close to the observations $\sim$ 130 days after explosion. However, the V- and I-band magnitudes show significant discrepancies (Fig.~\ref{fig:snlc_ubvri}) on the tail of the light curve, with the V-band overestimated by $\sim$ 1 magnitude and the I-band underestimated by $\sim$ 0.4 magnitude. The difference in the $^{56}\textrm{Ni}$ mass between the models and SN 1987A translates into $\sim$ 0.2 magnitude in all bands. This does not remove the discrepancy in the V and I bands. However, the overestimation in the V-band and the underestimation in I-band indicate that strong line-blanketing effects are missing in the models. These would redistribute the V-band fluxes to the I-band, possibly alleviating the discrepancy. Future studies with more complete model atoms may quantify such redistribution. In fact, the non-thermal sequence with Fe\,{\sc i} reduces the V magnitudes by $\sim$ 0.2 magnitude.

\begin{figure}
  \begin{center}
  \leavevmode
  \includegraphics[bb=50 270 595 520, width=0.48\textwidth]
  {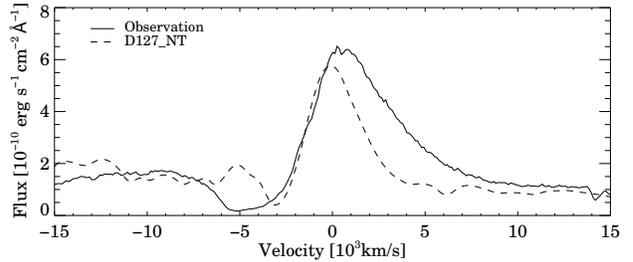}
  \caption{Same as Fig.~\ref{fig:cmp_snspec_3000_8000}, but it now shows
  the comparison of the H$\alpha$ profile in the velocity space between
  observation (solid) and model D127\_NT (dashed).}
  \label{fig:cmp_snspec_obs_ha}
  \end{center}
\end{figure}

\subsection{Spectral comparison}

Fig.~\ref{fig:cmp_snspec_3000_8000} shows a comparison of the observed optical spectra against the model spectra of D127\_NT (red), D127 (blue) and D127\_NT\_FeI (green). The H$\alpha$ profile is too weak in model D127, while both non-thermal models, D127\_NT and D127\_NT\_FeI, show considerably stronger H$\alpha$. Models D127\_NT and D127 show much stronger fluxes between 3000\,\AA\ and 6000\,\AA, which is reflected by the 1-magnitude offset in the V-band magnitudes in the previous section. The inclusion of Fe\,{\sc i} improves the fit, mainly in the wavelength range from 3000\,\AA\ to 4500\,\AA.

In Fig.~\ref{fig:cmp_snspec_obs_ha}, we show the theoretical and observed H$\alpha$ profiles in velocity space. While the peak of H$\alpha$ in the non-thermal model is comparable to that of the observation the profile is narrower. The left emission wing of the profile fits the observation, but the H$\alpha$ profile in the observation extends at least 3000~\kms\ further to the red than does the theoretical profile. Further, the P Cygni absorption component does not match --- the minimum of the model absorption sits at 3000~\kms, while observational minimum is at 5000~\kms.

\begin{figure*}
  \begin{center}
  \includegraphics[bb=30 215 590 575, width=0.98\textwidth]
  {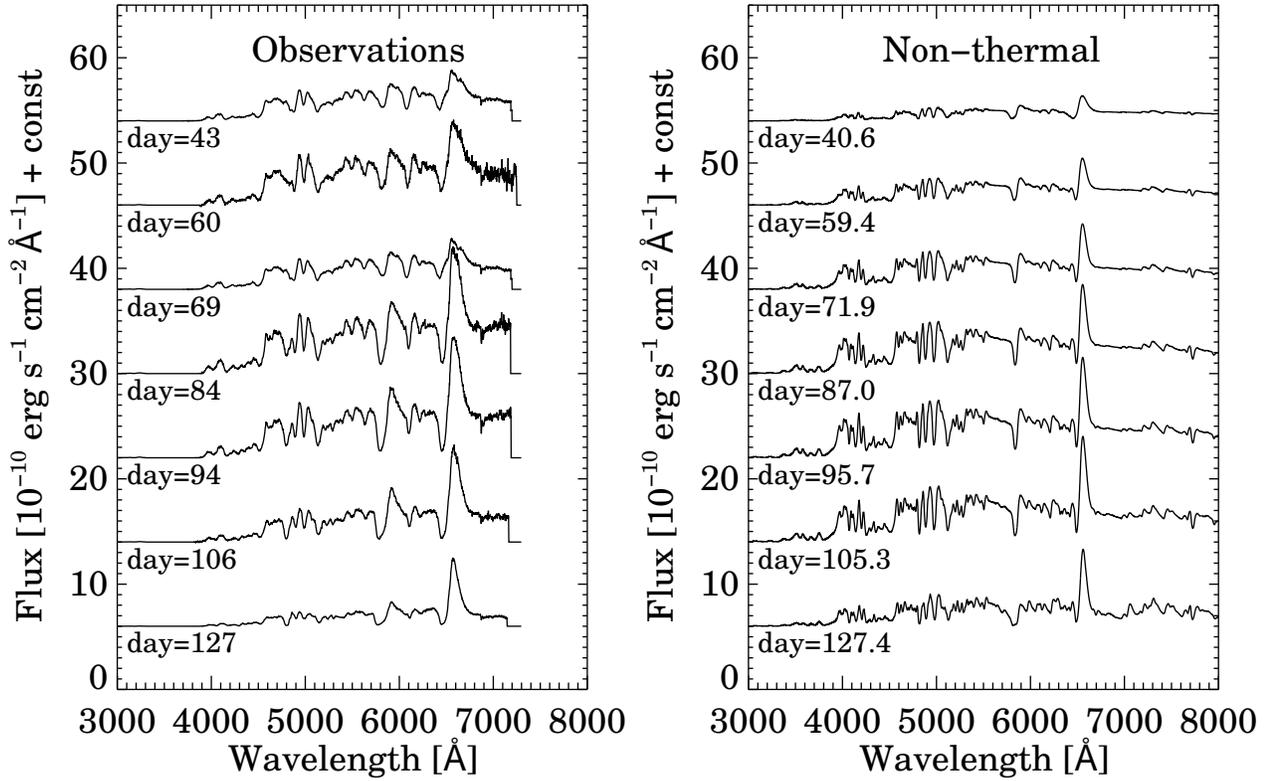}
  \caption{Left panel: montage of observed optical spectra of SN 1987A.
  The days since breakout are shown below each spectra. Right panel: montage
  of synthetic spectra of the non-thermal sequence. Each observed spectrum on
  the left is compared to a synthetic spectrum at roughly the same time since
  explosion. The synthetic spectra are reddened with
  E(B-V) = 0.15 and scaled for a distance of 50 kpc.}
  \label{fig:cmp_snspec_evol}
  \end{center}
\end{figure*}

The full width at half maximum (FWHM) of H$\alpha$ in the synthetic spectra is about 4000 \kms, indicating that the main region that contributes to H$\alpha$ is the core zone below 2000 \kms. This is coincident with the region in which $^{56}\textrm{Ni}$ is abundant. The spatial distribution of $^{56}\textrm{Ni}$ depends on the degree of mixing. Mixing effects in SNe explosion are widely seen in 2D and 3D models, however in pure 1D hydrodynamic models mixing must be induced artificially. According to the observed broad H$\alpha$ profile, our 1D hydrodynamical model has insufficient mixing of $^{56}\textrm{Ni}$. To constrain the mixing is not the goal of this paper, but the influence of mixing will be discussed in Section \ref{sec:discussion}. Another possible explanation is $\gamma$-ray transport. We made the assumption that all $\gamma$-rays deposit energy locally, but $\gamma$-ray transport makes it possible to deposit more energy at larger velocities. A discussion of the possible influence of $\gamma$-ray transport will be carried out in Section \ref{sec:discussion}.

\subsection{Comparison of the spectral evolution}
\label{sec:cmp_spec_evol}

To better compare the spectral features between the observations and the synthetic spectra, we selected the synthetic spectra which are about 10 days later than the observations. In Fig.~\ref{fig:cmp_snspec_evol}, the observations are shown on the left panel and the synthetic spectra are shown on the right panel. Recall that Balmer lines disappear suddenly at the end of the photospheric phase in the thermal sequence, thus the biggest improvement of the non-thermal sequence is the persistence of the strong H$\alpha$ profile at late times. In the non-thermal sequence, the strength of the H$\alpha$ profile first increases, and then decreases slowly, which is consistent with the observations. The Na\,{\sc i}\,D lines are contaminated by He\,{\sc i}\,5875~\AA. However, Na\,{\sc i}\,D lines show behavior similar to H$\alpha$, being narrower and weaker than the observations. This is also related to the non-thermal effects, since higher H ionization produces more electrons, which leads to an increase in neutral Na \citep{2008MNRAS.383...57D}. The synthetic spectra are rich with weak lines around 4000 \AA, while the observations show few weak lines there. Fe\,{\sc ii} is the main contributor to these lines. With the inclusion of Fe\,{\sc i}, this discrepancy becomes much smaller. Moreover, the model does not contain scandium and barium, which may explain a few missing P~Cygni profiles in the synthetic spectra. The absence of Ba\,{\sc ii} 6142~\AA\ is suggestive. The absence of Sc\,{\sc ii} in our model likely leads to an underestimate of line blanketing around 4000~\AA\ \citep{2011MNRAS.410.1739D}.

\section{UNCERTAINTIES}\label{sec:uncertainties}

\subsection{Impact excitation and ionization cross sections}
\label{sec:exc_ion_crosec}

\begin{figure*}
  \subfigure[Scale excitation cross sections]{\label{fig:cmp_exc_chanfrac}
  \includegraphics[bb=40 220 580 560, width=0.48\textwidth]
  {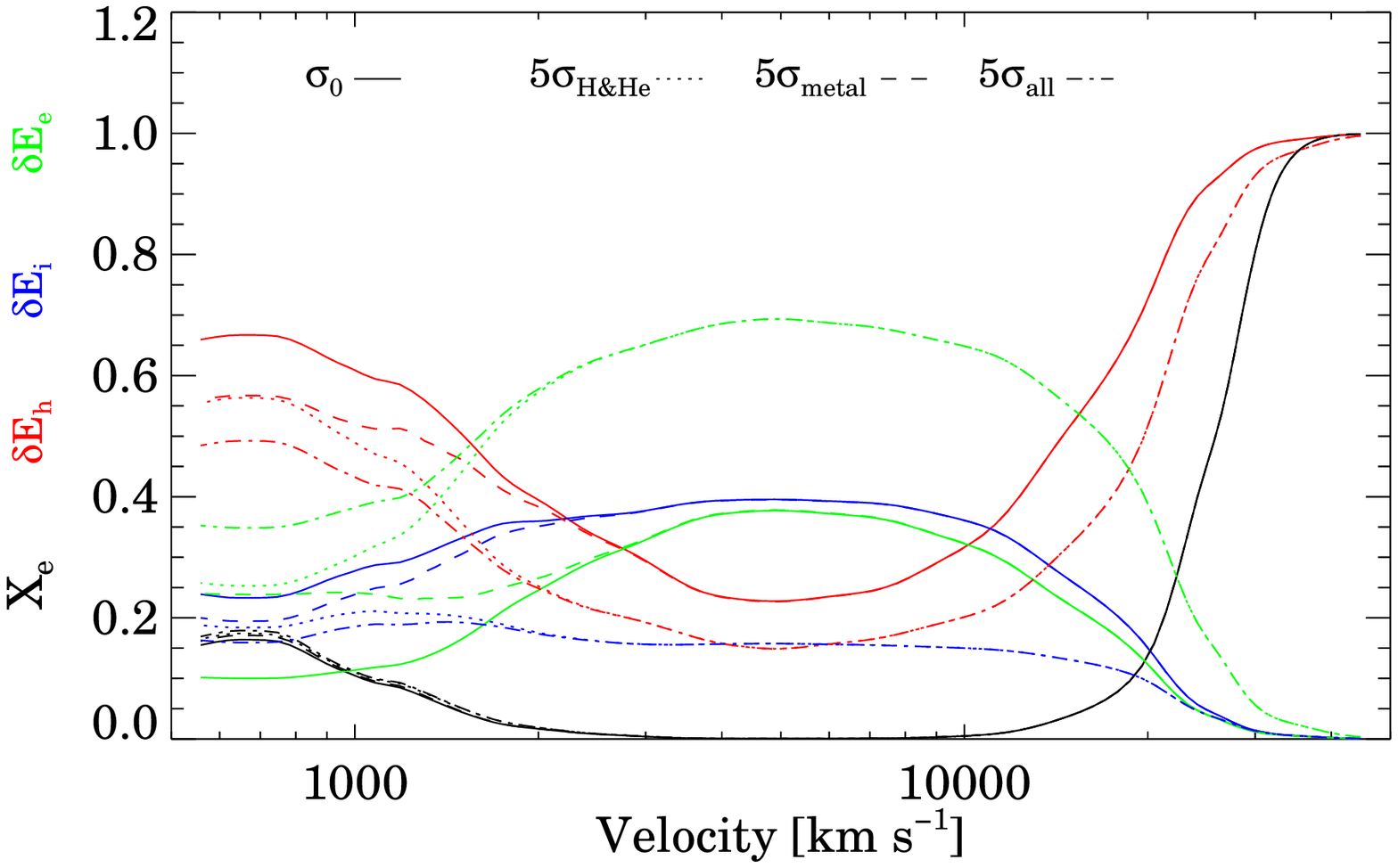}}
  \subfigure[Scale ionization cross sections]{\label{fig:cmp_ion_chanfrac}
  \includegraphics[bb=40 220 580 560, width=0.48\textwidth]
  {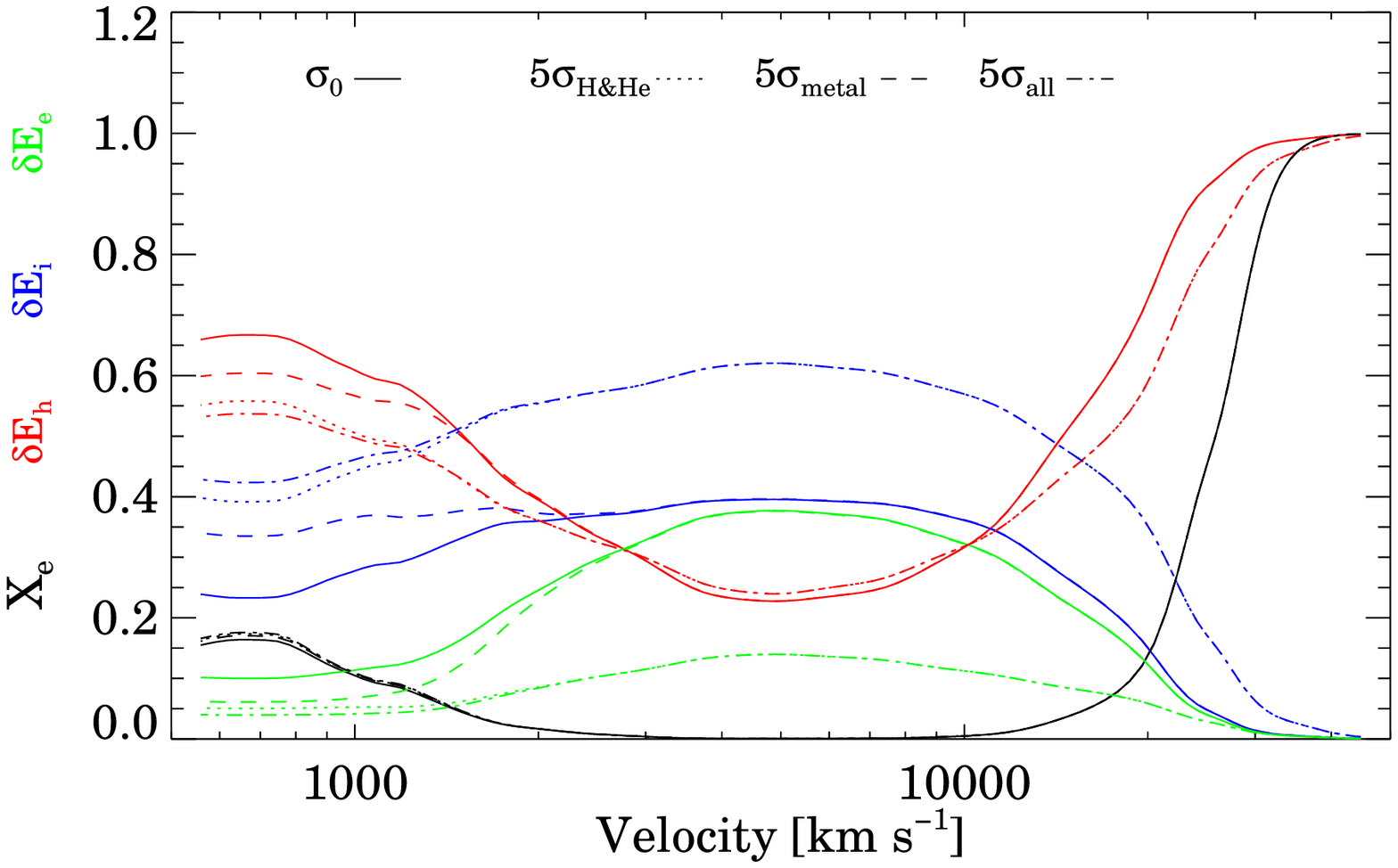}}
  \caption{Comparison of the energy fractions of the three $\gamma$-ray decay
  channels in models in which we have artificially scaled the excitation or ionization cross
  sections. The model used for testing is the one at day 127 without
  including Fe\,{\sc i}. Left: The 3 test cases of scaling the impact excitation
  cross sections. The scaling conditions are denoted in the figure.
  Different colors are fractions of different channels. The solid lines
  shows the fractions of the original model. The dotted lines, the dashed
  lines and the dashed-dotted lines represent the EH5 model, the EM5 model
  and the EA5 model, respectively. Right: The same as the left, but it
  shows the 3 cases of scaling the impact ionization cross sections. The
  dotted lines, the dashed lines and the dashed-dotted lines show the
  channel fractions of the IH5, IM5 and IA5 model, respectively.}
  \label{fig:cmp_chanfrac}
\end{figure*}

The uncertainty of the non-thermal solver mainly comes from the uncertainties of the electron impact excitation and ionization cross sections. As mentioned previously, the impact excitation cross sections are computed with the Bethe approximation, and the impact ionization cross sections utilize the analytical formula of \cite{1985A&AS...60..425A} with updated coefficients. The Bethe approximation is  a high-energy approximation derived from the Born approximation that neglects the short range interaction between the perturbing electron and the atomic electron. \cite{1962ApJ...136..906V} corrected the Gaunt factor
according to existing experimental data and accurate calculations to allow the formula to be used at low energies. The Bethe approximation generally works better at high energies. For some less understood species, the Bethe approximation may break down even at high energies. The \citeauthor{1985A&AS...60..425A} formula may have troubles for elements heavier than helium. Both experimental and theoretical cross sections are required to evaluate their uncertainties, but both of them are very difficult to estimate \citep{1994JPhB...27.5341B,2007A&A...475..765R,2009ADNDT..95..910R}.

\begin{table}
  \caption{Combinations of species to scale the cross sections}
  \label{tab:crosec_tests}
  \begin{center}
    \leavevmode
    \begin{tabular}{lll} \hline \hline
  Model            & excitation           &  ionization  \\ \hline
  EH5              & H and He by 5        &  Same   \\
  EM5              & Metal species by 5   &  Same   \\
  EA5              & All species by 5     &  Same   \\
  IH5              & Same                 &  H and He by 5        \\
  IM5              & Same                 &  Metal species by 5   \\
  IA5              & Same                 &  All species by 5     \\  \hline
    \end{tabular}
  \end{center}
\end{table}

To investigate the influence of the uncertainties, we ran six test models by artificially scaling the impact excitation or ionization cross sections for different combinations of species. These test models were all based on the same input model computed using standard cross-sections, and were run until  the populations and the radiation field were fully converged.
Table~\ref{tab:crosec_tests} lists the combinations and scalings applied to the cross sections. In Fig.~\ref{fig:cmp_chanfrac}, we compare the fractional energies of the three channels in these six tests. In the case of varying the excitation cross sections (Fig.~\ref{fig:cmp_chanfrac}, left panel), increasing the excitation cross sections of H and He results in an increase of fractional energy in the excitation channel and a decrease of fractional energy in the ionization channel. The change in the heating fraction is relatively small. An increase in the excitation cross sections of metal species has no effect above 2000~\kms, which is obviously due to their abundance deficiency. Below 2000~\kms, this also tends to increase the fraction of energy in the excitation channel and to decrease the fraction of energy in the ionization channel. The ejecta ionization fraction X$_e$ in all cases is roughly the same, implying that the ionization structure of the ejecta is hardly affected. Varying the ionization cross sections (Fig.~\ref{fig:cmp_chanfrac}, right panel) has the reverse effect of varying the excitation cross sections -- the increase of ionization cross sections results in an increase in the fractional energy of the ionization channel and a decrease in the fractional energy of the excitation energy. This is what we expect from the Spencer-Fano equation, given that the source term S(E) is fixed.

The fractional heating remains almost unchanged between 2500~\kms\ and 10\,000~\kms\ when the ionization cross sections increase. This surprising behavior is related to a low ejecta ionization fraction, X$_e$, seen in Fig.~\ref{fig:cmp_ion_chanfrac}. At these low-ionization fractions, the degradation spectra show that almost all electrons have degraded into very low energies and can no longer cause ionizations. Since an enhancement in the ionization cross-section does not dramatically change the number of secondary electrons produced, the ionization channel gains energy at the expense of the excitation channel. Conversely, a change in the excitation cross-sections does change the amount of energy entering the thermal channel. Crudely, excitation causes higher energy losses before the creation of a secondary electron via ionization. The competition between various processes is extremely sensitive to X$_e$. As X$_e$ becomes appreciable (e.g., above 10\,000 \kms.), many degraded electrons possess high energy. Increasing either the ionization or the excitation cross sections strengthens non-thermal processes at the expense of thermal energy deposition.

There are two types of uncertainties introduced by the approximations in computing the excitation and ionization cross sections. One is the uncertainty in the electron degradation spectrum. The second is the uncertainty in the rate for an individual process. For hydrogen these are explicitly coupled, but this is not necessarily so for other species since such species have a negligible influence on the degradation spectrum. In the six test cases, it turns out that the model spectra show no sizable difference. The H$\alpha$ profile shows a slightly stronger dependence on the impact cross sections of hydrogen. However, the two approximations adopted to compute the excitation and ionization cross sections produce fairly good calculations for hydrogen, which reduces our concern about such uncertainties. Generally, a factor of five in the uncertainties of the cross sections will not dramatically influence our results.

\subsection{E$_{\textrm{max}}$ for high energy electrons}

Radioactive decay from $^{56}\textrm{Ni}$ and $^{56}\textrm{Co}$ generally produces fast electrons with energy $\sim$ 1\,MeV. However, in the non-thermal solver, we assume a source function that injects all electrons with an energy E$_{\textrm{max}}$ = 1\,keV, which is very small compared to reality. While this high energy cut also introduces another source of uncertainty,
the  Bethe approximation, the Arnaud $\&$ Rothenflug formula, and the electron thermal term, have similar asymptotic behaviors at high energy, so the uncertainty should be small as long as E$_{\textrm{max}}$ is large enough. We have run two additional models with E$_{\textrm{max}}$ = 2\,keV and E$_{\textrm{max}}$ = 0.5\,keV. The difference in the fractional energies entering the three channels between one of the test models and the model with E$_{\textrm{max}}$ = 1\,keV is less than 5\%, and the difference between the two test models is slightly larger. Importantly, the choice of E$_{\textrm{max}}$ makes no difference to the spectra, despite the effects on the deposition fractions. Previous work by \citet{1991ApJ...375..190X} showed that the deposition fractions only have a weak sensitivity to E$_{\textrm{max}}$.

\subsection{The time-dependence effects on non-thermal processes}

Another uncertainty comes from the late inclusion of non-thermal excitation and ionization. At very early times the region where $^{56}\textrm{Ni}$ decays is in LTE, and the assumption that all the energy goes into heating is excellent. However, we included non-thermal excitation and ionization on day 40.6, when the time-dependent effect is already functioning at large velocities, in regions of low ionization and underabundant in $^{56}$Ni/$^{56}$Co. We study this uncertainty in two ways. The first method is to compare the thermal and the non-thermal model at day 40.6. At this epoch, non-thermal excitation and ionization are still unimportant. The model spectrum is hardly affected by the inclusion of non-thermal processes. This is also the main reason why we start the non-thermal sequence at this epoch. The second method is to compare two non-thermal models at a later time -- one is evolved from day 40.6 and the other with immediate inclusion of non-thermal processes. At day 127, we found very subtle differences between the spectra of the two models. The H$\alpha$ profile is only slightly weaker in the model with immediate inclusion of non-thermal processes. The above two tests indicate that the late inclusion of non-thermal excitation and ionization only introduces very small effects on the nebular spectra. Future studies will start with non-thermal processes, thereby removing such uncertainty.

\section{DISCUSSION}\label{sec:discussion}

The amount of mixing in SN 1987A has been widely debated (See \citeauthor{1989ARA&A..27..629A} \citeyear{1989ARA&A..27..629A} for a review). The two fundamental questions are what velocity hydrogen is mixed down to, and what velocity $^{56}\textrm{Ni}$ is mixed outward to. The latest 3D simulations of CCSNe \citep{2010ApJ...714.1371H} showed that significant amounts of hydrogen are transported into deep layers of the ejecta ($\le$ 1000 \kms), and that a large amount of heavy elements are strongly mixed above 2000~\kms. These results are thought to be very similar to the case of SN 1987A. However, it is difficult to draw a conclusion, since the simulations have a different progenitor from SN 1987A. \citet{1998ApJ...497..431K} modeled the late time spectra of SN 1987A and found hydrogen was mixed down to $\le$ 700~\kms (they needed the mixing to fit the observations).

\citet{1999MNRAS.302..314F} modeled the He\,{\sc i}\,1.083~$\mu$m line and found some amount of $^{56}\textrm{Ni}$ is required to be mixed above 3000~\kms\ to obtain a best fit. \citet{2001ApJ...556..979M} synthesized the spectrum of SN 1987A and argued that the strong Balmer lines at early times is due to non-thermal effects, which required a substantial amount of $^{56}\textrm{Ni}$ mixing out to 10\,000~\kms\ in the hydrogen envelope -- velocities much larger than predicted in hydrodynamic simulations. However, \citet{2005A&A...441..271U} and \cite{2008MNRAS.383...57D} showed time-dependent ionization is the main reason for the strong Balmer lines. The inward mixing of hydrogen and outward mixing of heavy elements in SN 1987A are still open questions.

The scale of mixing also plays an important role. Microscopic mixing results in homogeneous compositions in the ejecta, while macroscopic mixing can lead to large-scale compositional inhomogeneities. With microscopic mixing the mixed species can interact directly (e.g., charge exchange of H with Fe). With macroscopic mixing, this will not occur even if H and Fe are at the same radius (but different $\theta$, $\phi$). Microscopic mixing can be treated in 1D, while an accurate treatment of macroscopic mixing may require 3D models. In SN 1987A, the high velocity feature of iron-group emission lines (e.g., [Fe\,{\sc ii}]\,17.94\,$\mu$m and 25.99\,$\mu$m \citep{1990ApJ...360..257H}; [Ni\,{\sc ii}]\,6.6\,$\mu$m \citep{1994ApJ...427..874C}), at late times strongly indicate macroscopic mixing. \citet{1989ApJ...343..323F} showed that microscopic mixing can significantly change the observed spectra at late times, and \citet{1993ApJ...419..824L} emphasized the importance of a clumpy structure on the Fe/Co/Ni lines. \citet{2011A&A...530A..45J} were able to reproduce reasonably the SN 1987A spectra 8 years after explosion with a purely macroscopic mixing model. Furthermore, hydrodynamical models were unable to produce efficient microscopic mixing \citep{1990ApJ...360..242S,1991A&A...251..505M,1994ApJ...425..814H}.

Modeling the late time spectra of SN~1987A provides insight into the amount of mixing. The disappearance of Balmer continuum photons at the nebular phase makes non-thermal excitation and ionization the only way to reproduce the Balmer lines. However, the problem is complicated by the $\gamma$-ray transport. $\gamma$-rays interact with the medium through three processes: photoelectric ionization, Compton scattering and pair production. The decay of $^{56}\textrm{Ni}$ and $^{56}\textrm{Co}$ mainly produces $\gamma$-ray with energy of $\sim$ 1 MeV. At this energy range, Compton scattering is the dominant process. To give a rough estimate of the mean free path of a high energy electron before it is Compton scattered by atoms, we adopt an effective, purely absorptive, gray opacity $\kappa_{\gamma} = 0.06 Y_e$~cm$^2$\,g$^{-1}$ \citep{1995ApJ...446..766S}, where $Y_e$ is the total number of electrons per baryon. The optical depth is calculated by,

\begin{equation}
d\tau = - \kappa_{\gamma} \rho ds
\end{equation}

\noindent Assuming $Y_e$ = 0.5 and integrating $\tau$ from velocity 1500~\kms, the place where $^{56}\textrm{Co}$ is still abundant, we obtain the mean free path of $\gamma$-ray is $\sim 1.0 \times 10^{14}$~cm, which means the photon can travel from velocity 1500 to 1700~\kms\ in model D127\_NT. No doubt, very few $\gamma$-rays  travel a long way before they interact with the medium and deposit their energies. A Monte Carlo code has been developed for computing the $\gamma$-ray transport \citep{2012arXiv1204.0527H}, which will give a more detailed understanding of the mixing problem. We ran a test with allowance for $\gamma$-ray transport at day 127 and, as expected, there is only a slight enhancement in the line strength and width of H$\alpha$.

The first discovery of hard X-rays in SN~1987A was made by \citet{1987Natur.330..227S}, approximately 5 months after the explosion, and the first detection of $\gamma$-ray was made by \citet{1988Natur.331..416M}, shortly after the discovery of the hard X-rays. This suggests that $\gamma$-ray transport was beginning to become important at day 127. However, our model at day 154 gives an estimate that $\gamma$-rays can only travel from 1500 to 1800~\kms\ using the effective gray opacity $\kappa_{\gamma}$, which indicates an insufficient outward mixing of $^{56}\textrm{Ni}$ in the input. In fact, theoretical modeling of the bolometric, X-ray and $\gamma$-ray light curves require substantial outward mixing of $^{56}\textrm{Ni}$ into the hydrogen envelope \citep{1988A&A...197L...7K,1989ApJ...340..396A}.

\section{CONCLUSION}
\label{sec:conclusion}

We developed a non-thermal solver, which takes into account ionization and excitation due to non-thermal electrons created by $\gamma$-rays that arise from the decay of $^{56}\textrm{Ni}$ and its daughter products, and incorporated it into our fully non-LTE time-dependent code to model SNe. We use the model `lm18a7Ad' of Woosley on 0.27 day, enforced into homology as an initial input, and benchmark the non-thermal solver by comparing model results with the observations of SN 1987A.

Non-thermal models have lower temperature and more excited/ionized material in the region where the non-thermal processes are crucial. Fractional energy of non-thermal excitation/ionization is prominent at places with very small ejecta ionization fraction X$_e$. The spectral comparison with the thermal models shows that H$\alpha$ is strongly increased at nebular epochs. At late times, He\,{\sc i} lines, which are totally absent in the thermal models, are present in non-thermal models. While most He\,{\sc i} lines are significantly contaminated by other lines, He\,{\sc i}\,2.058\,$\mu$m provides an excellent opportunity to infer the influence of non-thermal processes on helium. He\,{\sc i}\,7065\AA\ is a possible optical line that can be used to infer the influence of non-thermal effects in the nebular epoch. There are many O\,{\sc i} lines locating near 1.129\,$\mu$m, and they are also strengthened by non-thermal processes. Most of other lines are only weakly affected at the epochs considered here.

Optical and IR He\,{\sc i} lines mainly originate below 2000~\kms. We confirm that non-thermal excitation is the most important process for He\,{\sc i} 2.058~$\mu$m. However, He\,{\sc i} 1.083~$\mu$m is due to cascades from higher levels, which indirectly relate to non-thermal ionization. Although photoionization and recombination are prominent processes for populating many levels, non-thermal excitation and ionization are the processes controlling the ionization balance.

We also compare the non-thermal models with the observations. The re-brightening of the bolometric light curve peaks about 10 days later in our models. An underestimate in the amount of outward mixing of $^{56}\textrm{Ni}$ is the main reason for the delay. Although the synthetic bolometric light curve agrees reasonably with the observations, multi-band light curves show discrepancies in various degrees and reveal potential differences between the hydrodynamical model and SN 1987A or possible problems in our modeling. The opacity issue, related to limited levels in the atomic data or some missing species, is one possibility, considering that the V-band is overestimated and the I-band is underestimated. H$\alpha$ maintains a strong profile at the nebular epochs in the non-thermal models, resembling observations of SN 1987A. However, the model H$\alpha$ profile is much narrower than what we observed in SN 1987A, probably due to insufficient outward mixing of $^{56}\textrm{Ni}$. Although hydrogen is abundant above 2000 \kms\ in the model, the lack of $^{56}\textrm{Ni}$ in this region and the assumption of local deposition of $\gamma$-rays make the non-thermal effects negligible there. We also compared the optical spectral evolution to the observations and found a satisfactory agreement.

The uncertainties introduced by various approximations and assumptions in the non-thermal solver have been investigated. The primary uncertainties are due to the not-well-known electron impact excitation and ionization cross sections of metal species. However, our results for a BSG model are largely dependent on accuracies of the cross sections for hydrogen and helium, since they are the most abundant two elements. The Bethe approximation and the \citeauthor{1985A&AS...60..425A} formula produce very good estimates of the excitation and ionization cross sections for these two elements. Therefore, our conclusions are not affected by these uncertainties. Another source of uncertainty comes from the computation of the degradation spectrum, which arises from the choice of the number of energy bins and the energy cutoff E$_{\textrm{max}}$ at the high end of the degradation spectrum. We demonstrate that our choice for the number of energy bins, N = 1000, is sufficient to give a reliable degradation spectrum and using E$_{\textrm{max}}$ = 1000\,eV in our model produces almost the same spectra as using E$_{\textrm{max}}$ = 2000\,eV. We also explore the time dependence of non-thermal processes and find there is only a very subtle effect on predicted spectra.

With the non-thermal solver, we are able to simulate SNe from photospheric to nebular phases continuously. The influence of non-thermal ionization and excitation on Type Ib and Type Ic models is being investigated \citep{2012inprep0000.0001D}. As more and more nebular spectra are available, the comparison of observations with models will allow us to place constraints on the hydrodynamic models and nucleosynthesis. The non-thermal solver also provides opportunities to constrain mixing effects in SNe.

\section*{Acknowledgements}
CL acknowledges support from NASA theory grant NNX10AC80G. DJH acknowledges support from STScI theory grant HST-AR-11756.01.A and NASA theory grant NNX10AC80G. LD acknowledges financial support from the European Community through an International Re-integration Grant, under grant number PIRG04-GA-2008-239184. We would also like to thank Bob Kurucz, the people involved with the NIST Atomic Data Base, and the participants of the Opacity and Iron Projects (with special thanks to Keith Butler, Sultana Nahar, Anil Pradhan and Peter Storey) for computing and supplying atomic data to the astrophysical community without their work these calculations would not be possible.

\bibliographystyle{mn2e}

\bibliography{CSonSNe}

\begin{thebibliography}{64}
\expandafter\ifx\csname natexlab\endcsname\relax\def\natexlab#1{#1}\fi

\bibitem[{{Arnaud} \& {Rothenflug}(1985)}]{1985A&AS...60..425A}
{Arnaud} M., {Rothenflug} R., 1985, \aaps, 60, 425

\bibitem[{{Arnett} {et~al}\mbox{.}(1989){Arnett}, {Bahcall}, {Kirshner}, \&
  {Woosley}}]{1989ARA&A..27..629A}
{Arnett} W.~D., {Bahcall} J.~N., {Kirshner} R.~P., {Woosley} S.~E., 1989,
  \araa, 27, 629

\bibitem[{{Arnett} \& {Fu}(1989)}]{1989ApJ...340..396A}
{Arnett} W.~D., {Fu} A., 1989, \apj, 340, 396

\bibitem[{{Bautista}(1997)}]{Bau97_FeI_phot}
{Bautista} M.~A., 1997, \aaps, 122, 167

\bibitem[{{Burke} {et~al}\mbox{.}(1994){Burke}, {Burke}, \&
  {Dunseath}}]{1994JPhB...27.5341B}
{Burke} P.~G., {Burke} V.~M., {Dunseath} K.~M., 1994, Journal of Physics B
  Atomic Molecular Physics, 27, 5341

\bibitem[{{Cardelli} {et~al}\mbox{.}(1989){Cardelli}, {Clayton}, \&
  {Mathis}}]{1989ApJ...345..245C}
{Cardelli} J.~A., {Clayton} G.~C., {Mathis} J.~S., 1989, \apj, 345, 245

\bibitem[{{Chugai}(1991)}]{1991SvAL...17..400C}
{Chugai} N.~N., 1991, Soviet Astronomy Letters, 17, 400

\bibitem[{{Colgan} {et~al}\mbox{.}(1994){Colgan}, {Haas}, {Erickson}, {Lord},
  \& {Hollenbach}}]{1994ApJ...427..874C}
{Colgan} S.~W.~J., {Haas} M.~R., {Erickson} E.~F., {Lord} S.~D., {Hollenbach}
  D.~J., 1994, \apj, 427, 874

\bibitem[{{Dessart} {et~al}\mbox{.}(2008){Dessart}, {Blondin}, {Brown},
  {Hicken}, {Hillier}, {Holland}, {Immler}, {Kirshner}, {Milne}, {Modjaz}, \&
  {Roming}}]{2008ApJ...675..644D}
{Dessart} L. {et~al.}, 2008, \apj, 675, 644

\bibitem[{{Dessart} \& {Hillier}(2005)}]{2005A&A...437..667D}
{Dessart} L., {Hillier} D.~J., 2005, \aap, 437, 667

\bibitem[{{Dessart} \& {Hillier}(2006)}]{2006A&A...447..691D}
{Dessart} L., {Hillier} D.~J., 2006, \aap, 447, 691

\bibitem[{{Dessart} \& {Hillier}(2008)}]{2008MNRAS.383...57D}
{Dessart} L., {Hillier} D.~J., 2008, \mnras, 383, 57

\bibitem[{{Dessart} \& {Hillier}(2010)}]{2010MNRAS.405.2141D}
{Dessart} L., {Hillier} D.~J., 2010, \mnras, 405, 2141

\bibitem[{{Dessart} \& {Hillier}(2011)}]{2011MNRAS.410.1739D}
{Dessart} L., {Hillier} D.~J., 2011, \mnras, 410, 1739

\bibitem[{{Dessart} {et~al}\mbox{.}(2012){Dessart}, {Hillier}, {Li}, \&
  {Woosley}}]{2012inprep0000.0001D}
{Dessart} L., {Hillier} D.~J., {Li} C., {Woosley} S., 2012, \mnras, (submitted)

\bibitem[{{Dessart} {et~al}\mbox{.}(2011){Dessart}, {Hillier}, {Livne}, {Yoon},
  {Woosley}, {Waldman}, \& {Langer}}]{2011MNRAS.414.2985D}
{Dessart} L., {Hillier} D.~J., {Livne} E., {Yoon} S.-C., {Woosley} S.,
  {Waldman} R., {Langer} N., 2011, \mnras, 414, 2985

\bibitem[{{Fassia} \& {Meikle}(1999)}]{1999MNRAS.302..314F}
{Fassia} A., {Meikle} W.~P.~S., 1999, \mnras, 302, 314

\bibitem[{{Fransson} \& {Chevalier}(1989)}]{1989ApJ...343..323F}
{Fransson} C., {Chevalier} R.~A., 1989, \apj, 343, 323

\bibitem[{{Gehrz} \& {Ney}(1990)}]{1990PNAS...87.4354G}
{Gehrz} R.~D., {Ney} E.~P., 1990, Proceedings of the National Academy of
  Science, 87, 4354

\bibitem[{{Haas} {et~al}\mbox{.}(1990){Haas}, {Erickson}, {Lord}, {Hollenbach},
  {Colgan}, \& {Burton}}]{1990ApJ...360..257H}
{Haas} M.~R., {Erickson} E.~F., {Lord} S.~D., {Hollenbach} D.~J., {Colgan}
  S.~W.~J., {Burton} M.~G., 1990, \apj, 360, 257

\bibitem[{{Hammer} {et~al}\mbox{.}(2010){Hammer}, {Janka}, \&
  {M{\"u}ller}}]{2010ApJ...714.1371H}
{Hammer} N.~J., {Janka} H.-T., {M{\"u}ller} E., 2010, \apj, 714, 1371

\bibitem[{{Harkness} {et~al}\mbox{.}(1987){Harkness}, {Wheeler}, {Margon},
  {Downes}, {Kirshner}, {Uomoto}, {Barker}, {Cochran}, {Dinerstein}, {Garnett},
  \& {Levreault}}]{1987ApJ...317..355H}
{Harkness} R.~P. {et~al.}, 1987, \apj, 317, 355

\bibitem[{{Herant} \& {Woosley}(1994)}]{1994ApJ...425..814H}
{Herant} M., {Woosley} S.~E., 1994, \apj, 425, 814

\bibitem[{{Hillebrandt} {et~al}\mbox{.}(1987){Hillebrandt}, {Hoeflich},
  {Weiss}, \& {Truran}}]{1987Natur.327..597H}
{Hillebrandt} W., {Hoeflich} P., {Weiss} A., {Truran} J.~W., 1987, \nat, 327,
  597

\bibitem[{{Hillier}(1987)}]{1987ApJS...63..947H}
{Hillier} D.~J., 1987, \apjs, 63, 947

\bibitem[{{Hillier} \& {Dessart}(2012)}]{2012arXiv1204.0527H}
{Hillier} D.~J., {Dessart} L., 2012, \mnras, (in press)

\bibitem[{{Hillier} \& {Miller}(1998)}]{1998ApJ...496..407H}
{Hillier} D.~J., {Miller} D.~L., 1998, \apj, 496, 407

\bibitem[{{Jerkstrand} {et~al}\mbox{.}(2011){Jerkstrand}, {Fransson}, \&
  {Kozma}}]{2011A&A...530A..45J}
{Jerkstrand} A., {Fransson} C., {Kozma} C., 2011, \aap, 530, A45

\bibitem[{{Kozma} \& {Fransson}(1992)}]{1992ApJ...390..602K}
{Kozma} C., {Fransson} C., 1992, \apj, 390, 602

\bibitem[{{Kozma} \& {Fransson}(1998{\natexlab{a}})}]{1998ApJ...496..946K}
{Kozma} C., {Fransson} C., 1998{\natexlab{a}}, \apj, 496, 946

\bibitem[{{Kozma} \& {Fransson}(1998{\natexlab{b}})}]{1998ApJ...497..431K}
{Kozma} C., {Fransson} C., 1998{\natexlab{b}}, \apj, 497, 431

\bibitem[{{Kumagai} {et~al}\mbox{.}(1988){Kumagai}, {Shigeyama}, {Nomoto},
  {Itoh}, \& {Nishimura}}]{1988A&A...197L...7K}
{Kumagai} S., {Shigeyama} T., {Nomoto} K., {Itoh} M., {Nishimura} J., 1988,
  \aap, 197, L7

\bibitem[{{Kurucz}(2009)}]{2009AIPC.1171...43K}
{Kurucz} R.~L., 2009, in American Institute of Physics Conference Series, Vol.
  1171, Recent Directions in Astrophysical Quantitative Spectroscopy and
  Radiation Hydrodynamics, {Hubeny} I., {Stone} J.~M., {MacGregor} K., {Werner}
  K., eds., pp. 43--51

\bibitem[{{Landolt}(1992)}]{1992AJ....104..340L}
{Landolt} A.~U., 1992, \aj, 104, 340

\bibitem[{{Langer} {et~al}\mbox{.}(1989){Langer}, {El Eid}, \&
  {Baraffe}}]{1989A&A...224L..17L}
{Langer} N., {El Eid} M.~F., {Baraffe} I., 1989, \aap, 224, L17

\bibitem[{{Li} \& {McCray}(1995)}]{1995ApJ...441..821L}
{Li} H., {McCray} R., 1995, \apj, 441, 821

\bibitem[{{Li} {et~al}\mbox{.}(1993){Li}, {McCray}, \&
  {Sunyaev}}]{1993ApJ...419..824L}
{Li} H., {McCray} R., {Sunyaev} R.~A., 1993, \apj, 419, 824

\bibitem[{{Lucy}(1991)}]{1991ApJ...383..308L}
{Lucy} L.~B., 1991, \apj, 383, 308

\bibitem[{{Matz} {et~al}\mbox{.}(1988){Matz}, {Share}, {Leising}, {Chupp}, \&
  {Vestrand}}]{1988Natur.331..416M}
{Matz} S.~M., {Share} G.~H., {Leising} M.~D., {Chupp} E.~L., {Vestrand} W.~T.,
  1988, \nat, 331, 416

\bibitem[{{Meikle} {et~al}\mbox{.}(1989){Meikle}, {Spyromilio}, {Varani}, \&
  {Allen}}]{1989MNRAS.238..193M}
{Meikle} W.~P.~S., {Spyromilio} J., {Varani} G.-F., {Allen} D.~A., 1989,
  \mnras, 238, 193

\bibitem[{{Mitchell} {et~al}\mbox{.}(2001){Mitchell}, {Baron}, {Branch},
  {Lundqvist}, {Blinnikov}, {Hauschildt}, \& {Pun}}]{2001ApJ...556..979M}
{Mitchell} R.~C., {Baron} E., {Branch} D., {Lundqvist} P., {Blinnikov} S.,
  {Hauschildt} P.~H., {Pun} C.~S.~J., 2001, \apj, 556, 979

\bibitem[{{Mueller} {et~al}\mbox{.}(1991){Mueller}, {Fryxell}, \&
  {Arnett}}]{1991A&A...251..505M}
{Mueller} E., {Fryxell} B., {Arnett} D., 1991, \aap, 251, 505

\bibitem[{{Opal} {et~al}\mbox{.}(1971){Opal}, {Peterson}, \&
  {Beaty}}]{1971JChPh..55.4100O}
{Opal} C.~B., {Peterson} W.~K., {Beaty} E.~C., 1971, \jcp, 55, 4100

\bibitem[{{Panagia}(2005)}]{2005coex.conf..585P}
{Panagia} N., 2005, in IAU Colloq. 192: Cosmic Explosions, On the 10th
  Anniversary of SN1993J, {J.-M.~Marcaide \& K.~W.~Weiler}, ed., pp. 585--+

\bibitem[{{Pelan} \& {Berrington}(1997)}]{PB97_FeI_col}
{Pelan} J., {Berrington} K.~A., 1997, \aaps, 122, 177

\bibitem[{{Phillips} \& {Heathcote}(1989)}]{1989PASP..101..137P}
{Phillips} M.~M., {Heathcote} S.~R., 1989, \pasp, 101, 137

\bibitem[{{Phillips} {et~al}\mbox{.}(1988){Phillips}, {Heathcote}, {Hamuy}, \&
  {Navarrete}}]{1988AJ.....95.1087P}
{Phillips} M.~M., {Heathcote} S.~R., {Hamuy} M., {Navarrete} M., 1988, \aj, 95,
  1087

\bibitem[{{Podsiadlowski} {et~al}\mbox{.}(1990){Podsiadlowski}, {Joss}, \&
  {Rappaport}}]{1990A&A...227L...9P}
{Podsiadlowski} P., {Joss} P.~C., {Rappaport} S., 1990, \aap, 227, L9

\bibitem[{{Ramsbottom}(2009)}]{2009ADNDT..95..910R}
{Ramsbottom} C.~A., 2009, Atomic Data and Nuclear Data Tables, 95, 910

\bibitem[{{Ramsbottom} {et~al}\mbox{.}(2007){Ramsbottom}, {Hudson},
  {Norrington}, \& {Scott}}]{2007A&A...475..765R}
{Ramsbottom} C.~A., {Hudson} C.~E., {Norrington} P.~H., {Scott} M.~P., 2007,
  \aap, 475, 765

\bibitem[{{Shigeyama} \& {Nomoto}(1990)}]{1990ApJ...360..242S}
{Shigeyama} T., {Nomoto} K., 1990, \apj, 360, 242

\bibitem[{{Shull} \& {van Steenberg}(1985)}]{1985ApJ...298..268S}
{Shull} J.~M., {van Steenberg} M.~E., 1985, \apj, 298, 268

\bibitem[{{Spencer} \& {Fano}(1954)}]{1954PhRv...93.1172S}
{Spencer} L.~V., {Fano} U., 1954, Physical Review, 93, 1172

\bibitem[{{Suntzeff} \& {Bouchet}(1990)}]{1990AJ.....99..650S}
{Suntzeff} N.~B., {Bouchet} P., 1990, \aj, 99, 650

\bibitem[{{Sunyaev} {et~al}\mbox{.}(1987){Sunyaev}, {Kaniovsky}, {Efremov},
  {Gilfanov}, {Churazov}, {Grebenev}, {Kuznetsov}, {Melioranskiy},
  {Yamburenko}, {Yunin}, {Stepanov}, {Chulkov}, {Pappe}, {Boyarskiy},
  {Gavrilova}, {Loznikov}, {Prudkoglyad}, {Rodin}, {Reppin}, {Pietsch},
  {Engelhauser}, {Truemper}, {Voges}, {Kendziorra}, {Bezler}, {Staubert},
  {Brinkman}, {Heise}, {Mels}, {Jager}, {Skinner}, {Al-Emam}, {Patterson},
  {Willmore}, {Gilfanov}, \& {Churazov}}]{1987Natur.330..227S}
{Sunyaev} R. {et~al.}, 1987, \nat, 330, 227

\bibitem[{{Swartz}(1991)}]{1991ApJ...373..604S}
{Swartz} D.~A., 1991, \apj, 373, 604

\bibitem[{{Swartz} {et~al}\mbox{.}(1993){Swartz}, {Filippenko}, {Nomoto}, \&
  {Wheeler}}]{1993ApJ...411..313S}
{Swartz} D.~A., {Filippenko} A.~V., {Nomoto} K., {Wheeler} J.~C., 1993, \apj,
  411, 313

\bibitem[{{Swartz} {et~al}\mbox{.}(1995){Swartz}, {Sutherland}, \&
  {Harkness}}]{1995ApJ...446..766S}
{Swartz} D.~A., {Sutherland} P.~G., {Harkness} R.~P., 1995, \apj, 446, 766

\bibitem[{{Utrobin}(2004)}]{2004AstL...30..293U}
{Utrobin} V.~P., 2004, Astronomy Letters, 30, 293

\bibitem[{{Utrobin} \& {Chugai}(2005)}]{2005A&A...441..271U}
{Utrobin} V.~P., {Chugai} N.~N., 2005, \aap, 441, 271

\bibitem[{{van Regemorter}(1962)}]{1962ApJ...136..906V}
{van Regemorter} H., 1962, \apj, 136, 906

\bibitem[{{Weaver} {et~al}\mbox{.}(1978){Weaver}, {Zimmerman}, \&
  {Woosley}}]{1978ApJ...225.1021W}
{Weaver} T.~A., {Zimmerman} G.~B., {Woosley} S.~E., 1978, \apj, 225, 1021

\bibitem[{{Xu} \& {McCray}(1991)}]{1991ApJ...375..190X}
{Xu} Y., {McCray} R., 1991, \apj, 375, 190

\bibitem[{{Younger}(1981)}]{1981JQSRT..26..329Y}
{Younger} S.~M., 1981, \jqsrt, 26, 329

\end{thebibliography}

\label{lastpage}

\end{document}